\documentclass[sigconf]{acmart}

\usepackage{graphicx} 
\usepackage{subfig} 

\usepackage{enumitem}
\usepackage{verbatim}
\usepackage{amsmath} 
\usepackage{algorithmicx}
\usepackage{algpseudocode}

\usepackage{xcolor}
\usepackage{array}
\usepackage{balance}
\usepackage{fancyvrb}
\usepackage{verbatim}
\usepackage{listings}
\usepackage{textcomp}
\usepackage{booktabs}
\usepackage{multirow}
\usepackage{pifont}
\usepackage{url}
\usepackage{tcolorbox}
\usepackage{listing}
\usepackage{mwe}
\usepackage{pgfplots}
\usepackage{filecontents}
\usepackage{tikz}
\usepackage{tikzscale}

\newcommand\markup[1]{{\color{red}#1}} 
\newcommand{\cmark}{\ding{51}}%

\newcommand{\codefont}[1]{\footnotesize{\texttt{#1}}\normalsize}
\newcommand{\totalcount}{497}
\newcommand{\implcount}{472}
\newcommand\measureISpecification{4ex}

\setcopyright{rightsretained}

\acmDOI{10.475/123_4}

\acmISBN{123-4567-24-567/08/06}

\begin{document}
\title{Secure Coding Practices in Java: Challenges and Vulnerabilities}

\author{Na Meng, Stefan Nagy, Daphne Yao, Wenjie Zhuang, Gustavo Arango Argoty}
\orcid{1234-5678-9012}
\affiliation{%
  \institution{Virginia Tech}
  \city{Blacksburg} 
  \state{Virginia} 
  \postcode{24060}
}
\email{{nm8247, snagy2, danfeng, kaito, gustavo1}@vt.edu}

\begin{abstract}
Java platform and third-party libraries provide various security features to facilitate secure coding.
However, misusing these features can cost tremendous time and effort of developers or cause security vulnerabilities in software. 
Prior research was focused on the misuse of cryptography and SSL APIs, but did not explore the key fundamental research question: what are the biggest challenges and vulnerabilities in secure coding practices?
In this paper, we conducted a comprehensive empirical study on StackOverflow posts to understand developers' concerns on Java secure coding, their programming obstacles, and potential vulnerabilities in their code.


We observed that developers have shifted their effort to the usage of authentication and authorization features provided by Spring security---a third-party framework designed to secure enterprise applications. 
Multiple programming challenges are related to APIs or libraries, including the complicated cross-language data handling of cryptography APIs, and the complex Java-based or XML-based approaches to configure Spring security.
More interestingly, \emph{we identified security vulnerabilities in the suggested code of accepted answers}. The vulnerabilities included using insecure hash functions such as MD5, breaking SSL/TLS security through bypassing certificate validation, and insecurely disabling the default protection against Cross Site Request Forgery (CSRF) attacks. Our findings reveal 
the insufficiency of secure coding assistance and education, and the gap between security theory and coding practices. 


\end{abstract}

%
%

\ccsdesc[500]{General and reference~Empirical studies}

\keywords{CSRF, SSL/TLS certificate validation, cryptographic hash functions, authentication, authorization}

\maketitle

\section{Introduction}
\label{sec:intro}
Java platform and third-party libraries or frameworks (e.g., BouncyCastle~\cite{bc} and Spring Security~\cite{springsecurity}) provide various features to facilitate secure coding.
However, misusing these libraries and frameworks not only costs excessive debugging effort of developers, but also leads to security vulnerabilities in software~\cite{libmisuse,cwe227,Shuai:2014,veracode}. For example, Veracode identified software errors in the handling of user credentials, including hard-coded password and plaintext passwords in configuration files~\cite{veracode}. These errors can enable attackers to bypass access controls.


Prior research mainly focused on the misuse of cryptography and SSL APIs that causes security vulnerabilities~\cite{Fahl:2012,Georgiev:2012,Egele:2013,Lazar:2014}. Specifically, Lazar et al.~manually examined 269 published cryptographic vulnerabilities in the CVE database, and observed 83\% of them were caused by cryptography API misuse~\cite{Lazar:2014}. 
Fahl et al.~\cite{Fahl:2012} and Georgiev et al.~\cite{Georgiev:2012} separately implemented the man-in-the-middle attack, and detected vulnerable Android applications and software libraries that misused SSL APIs.
Nadi et al.~further investigated the obstacles developers face while using the Java cryptography APIs, the tasks for which they use the APIs, and the kind of tool support they desire~\cite{Nadi:2016}. 
Despite these studies, some key questions on secure coding practices remain unanswered. They include (1) whether programmers are equipped with sufficient security knowledge and automatic coding support, and (2) whether the coding practices benefitted from security research over the years.


For this paper, we conducted \emph{a comprehensive in-depth investigation on the common concerns, programming challenges, and security vulnerabilities in developers' secure coding practices} by manually inspecting \totalcount{} StackOverflow posts related to Java security. We chose StackOverflow~\cite{stackoverflow} because (1) developers usually share and discuss programming issues and solutions on this online platform, and (2) StackOverflow plays an important role in educating developers and impacting their daily coding practices. The main challenge of performing this empirical study is \emph{interpreting each security-relevant programming issue or solution within both the program context and security context}. To comprehend each post within the program context, we manually checked all mentioned information about the source code, configuration files, and/or execution environments. Then we decided the root cause and solution of the problem. To comprehend each post within the security context, we also identified the security requirement that developers tried to implement and investigated the involved security libraries. Then we determined whether the implementation fulfilled the requirement. Such manual analysis requires so much comprehension and expertise in both software engineering (SE) and security that it is difficult to automate the process. 

With our thorough manual analysis on the 497 posts, we investigated the following three research questions (RQs):


\begin{itemize}
\item[\textbf{RQ1}] \emph{What are the common concerns on Java secure coding?}  
Although there are various security libraries and frameworks~\cite{Oaks:1998,Gong:2003,springsecurityreference32,asref,jaasref,jca},
several questions are still unanswered, such as (1) which are the popular security features being frequently asked about, and (2) what 
are the hard-to-implement security defenses in practice?
\item[\textbf{RQ2}] \emph{What are the common programming challenges?} 
We aim to identify the common obstacles that prevented developers from implementing secure code easily and correctly. This information will help guide SE researchers and tool builders to better develop tools, and to help close the gap between the intended library usage and developers' actual usage.
\item[\textbf{RQ3}] \emph{What are the common security vulnerabilities?}
We aim to identify security vulnerabilities in StackOverflow posts, because the bad practices recommended on the platform can become popular and cause profound negative impact. 
This effort will help raise the security consciousness of secure software practitioners.

\end{itemize}
In our study, we made three major observations.
\begin{itemize}
\item \emph{There were multiple security vulnerabilities in the recommended code of some accepted answers}. For instance, 
the usage of MD5 and SHA-1 algorithms were repetitively suggested, although these algorithms are notoriously insecure and should not be used anymore. Additionally, developers were advised to trust all incoming SSL/TLS certificates from servers as a workaround to certificate verification errors.
Such practice completely disables the security checks of SSL.
Although this bad practice was initially reported by researchers in 2012~\cite{Fahl:2012,Georgiev:2012}, developers have still asked for and accepted the practice till now.
Furthermore, when implementing authentication with Spring security and getting errors, developers were suggested with a workaround solution to blindly disable the default security protection against Cross Site Request Forgery (CSRF) attacks.
\item \emph{There were various programming challenges related to security libraries}. For instance, developers were stuck with cryptography API usage due to clueless error messages, complex cross-language data handling, and delicate implicit API usage constraints.
However, when using Spring security, developers struggled a lot with the two alternative ways of configuring security: Java-based or XML-based.
\item \emph{Software developers have shifted their security implementation effort to Spring security since 2012.} 261 of the 497 examined posts (53\%) were about Spring security.
However, we have not seen any research that checks or analyzes the security vulnerabilities related to the framework.
\end{itemize}

The significance of this work is that we provided empirical evidence for a significant number of alarming secure coding issues, which have not been previously reported. These issues are due to a variety of reasons, including the rapidly increasing need for enterprise security applications, the lack of  security training in the software development workforce, and poorly designed security libraries.  We hope our findings can motivate the community to research solutions for helping developers overcome these obstacles in the long term.



\section{Background}
\label{sec:background}

The examined StackOverflow posts were mainly about three perspectives of Java security: 
Java platform security, Java EE security, and other third-party frameworks. This section introduces the key terminologies used throughout the paper.

\subsection{Java Platform Security}
The platform defines APIs spanning major security areas, including cryptography, access control, and secure communication~\cite{jSecurity}.

\emph{The Java Cryptography Architecture (JCA)} contains 
APIs for 
\textbf{hashes}, \textbf{keys and certificates}, \textbf{digital signatures}, and \textbf{encryption}~\cite{jca}. 
Nine cryptographic engines are defined to provide either
cryptographic operations (encryption, digital signatures, hashes), 
generators or converters of cryptographic material (keys and algorithm parameters), or
objects (keystores or certificates) that encapsulate the cryptographic data.

\emph{The access control architecture} protects the access to sensitive resources (e.g., local files) or sensitive application code (e.g., methods in a class). All access control decisions are mediated by a \textbf{security manager}. 
By default, the security manager uses the \codefont{AccessController} class for access control operations and decisions. 

\emph{Secure communication} ensures that the data which travels across a network is sent to the appropriate party, without being modified during the transmission. Cryptography forms the basis for secure communication. The Java platform provides API support for standard secure communication protocols like \textbf{SSL/TLS}.
\textbf{HTTPS}, or ``HTTP secure'', is an application-specific implementation that is a combination of HTTP and SSL/TLS. 


\subsection{Java EE Security}
Java EE is an standard specification for enterprise Java extensions~\cite{jeesecurity}. 
Various application servers are built to implement this specification, such as \textbf{JBoss} or \textbf{WildFly}~\cite{wildfly}, \textbf{Glassfish}~\cite{glassfish}, \textbf{WebSphere}~\cite{websphere}, and \textbf{WebLogic}~\cite{weblogic}.
A Java EE application consists of components deployed into various containers. 
The Java EE security specification defines that containers secure components by supporting features like authentication and authorization. 

In particular, \textbf{authentication} defines how communicating entities, such as a client and a server, prove to each other that they are 
who they say they are.
An authenticated user is issued a \emph{credential}, which includes user information like usernames/passwords or tokens.
\textbf{Authorization} 
ensures that users have permissions to perform operations or access data. When accessing certain resource, a user is authorized if the server can map this user to a security role permitted for the resource.

Security for Java EE applications can be implemented in the following two ways:
\begin{itemize}
\item \textbf{Declarative Security} expresses an application component's security requirements using either \textbf{deployment descriptors} or \textbf{annotations}. A deployment descriptor is an XML file external to the application. This XML file expresses an application's security structure, including security roles, access control, and authentication requirements. Annotations are used to specify security information in a class file. They can be either used by or overridden by deployment descriptors.
\item \textbf{Programmatic Security} is embedded in an application and is used to make security decisions, when declarative security alone is not sufficient to express the security model.
\end{itemize}

\subsection{Other Third-Party Frameworks}
Several frameworks were built to provide authentication, authorization, and other security features for enterprise applications, such as \textbf{Spring Security}~\cite{springsecurity}.
Different from the Java EE security APIs, these frameworks are container independent, meaning that they do not require containers to implement security. For example, Spring security is installed as a single \textbf{Filter} in the filter chain inside a container to handle requests.
There can be multiple security filters inside Spring security.
Developers can configure security in an \textbf{XML-based} way, a \textbf{Java-based} way, or a hybrid of the two. Similar to Java EE security, the XML-based way implements security requirements with deployment descriptors and source code, while the Java-based way expresses security with annotations and code.

\section{Methodology}
\label{sec:method}
\begin{figure}[!hbt]
\centering
\scalebox{0.41}{\includegraphics{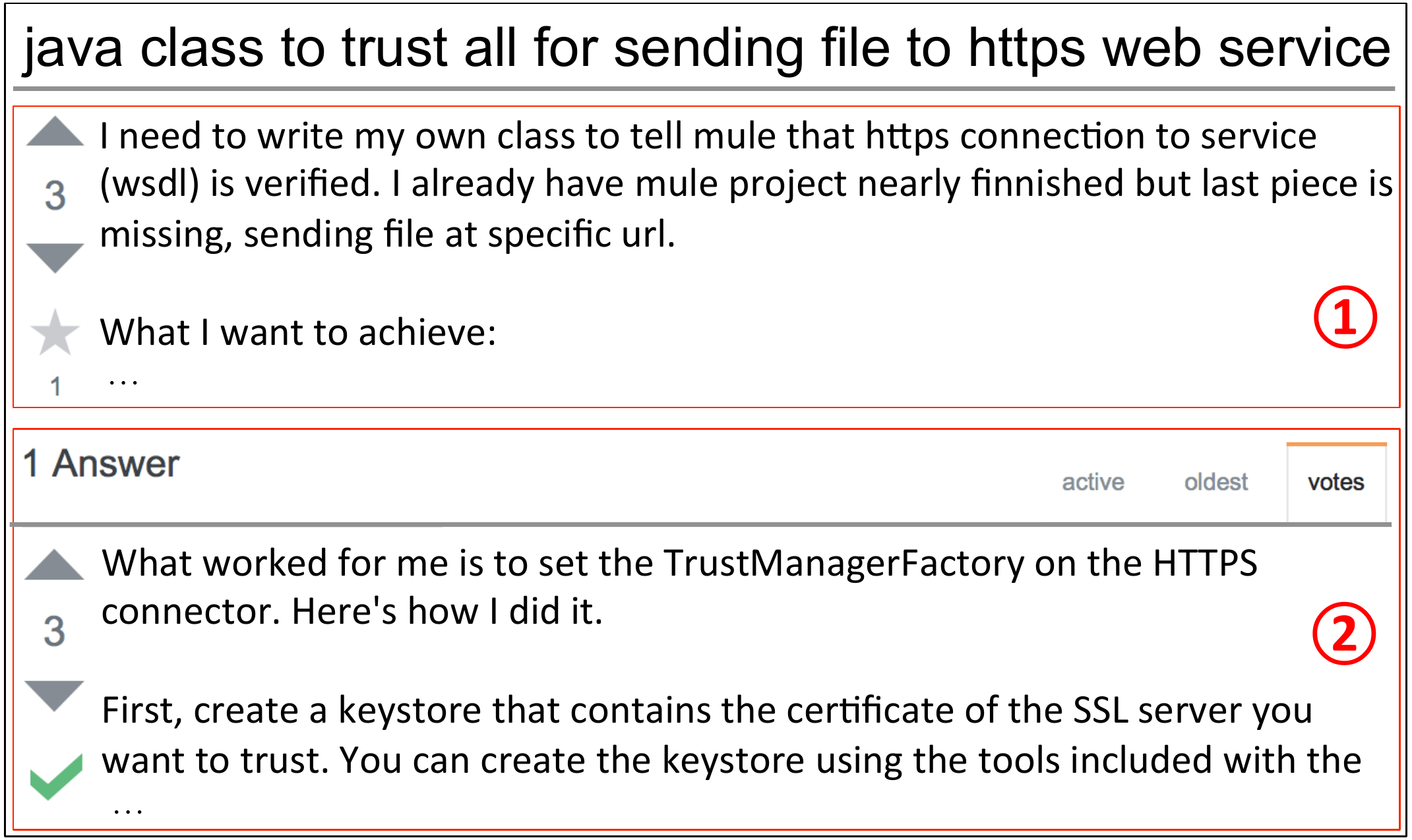}}
\vspace{-2em}
\caption{A highly viewed post asking about workarounds to bypass key checking and allow all host names for HTTPS~\cite{post}}
\label{fig:post}
\end{figure}

We leveraged Scrapy---an open source python library~\cite{scrapy} to crawl posts from the StackOverflow website. 
Figure~\ref{fig:post} presents the format of a typical StackOverflow post. Each post mainly contains two regions: the question and answers. 

\textcircled{1} \textbf{Question region} contains the question description and some metadata.
The metadata includes a \textbf{vote} for the question (e.g., 3)---indicating whether the question is well-defined or well-representative, and a 
\textbf{favorite count} (e.g., 1)---showing how many people liking the question.

\textcircled{2} \textbf{Answer region} contains all answer(s) provided. When one or more answers are provided, the asker decides which answer to \textbf{accept}, and marks it with (\textcolor{green}{\cmark}). 

\vspace{-0.5em}
We obtained 22,195 posts containing keywords ``java'' and ``security''. 
After extracting the question, the answers, and relevant metadata for each post, we refined the data in three ways. 

\emph{1) Filtering less useful posts.} We automatically refined posts by removing duplicated posts, posts without accepted answers, and posts whose questions received negative votes perhaps because the questions were ill-formed or confusing.

\emph{2) Removing posts without code snippets.} We only focused on posts containing code snippets to better understand the questions within the program context.
Since our crawled data did not include any metadata to describe the existence of code snippets, we developed an intuitive filter to search for keywords ``public'' and ``class'' in each post. Based on our observation, a post usually contains these two keywords when it includes a code snippet. 

\emph{3) Discarding irrelevant posts.}
After applying the above two filters, we manually examined the remaining posts, and decided whether they were relevant to Java secure coding or simply contained the checked keywords accidentally.

With the above three filters, 
we finally included \totalcount{} posts in our data set. The question asking time of these posts were during 2008-2016. We did not include the posts in 2017, because at the time we conducted experiments, there was only data for the first several months of 2017.
When manually filtering retrieved posts, we also characterized relevant posts based on their \emph{security concerns}, \emph{programming challenges}, and \emph{security vulnerabilities}. Based on the characterization, we classified posts and investigated the following three research questions (RQs):

\begin{figure*}[!htb]
\centering
\includegraphics[width=11cm]{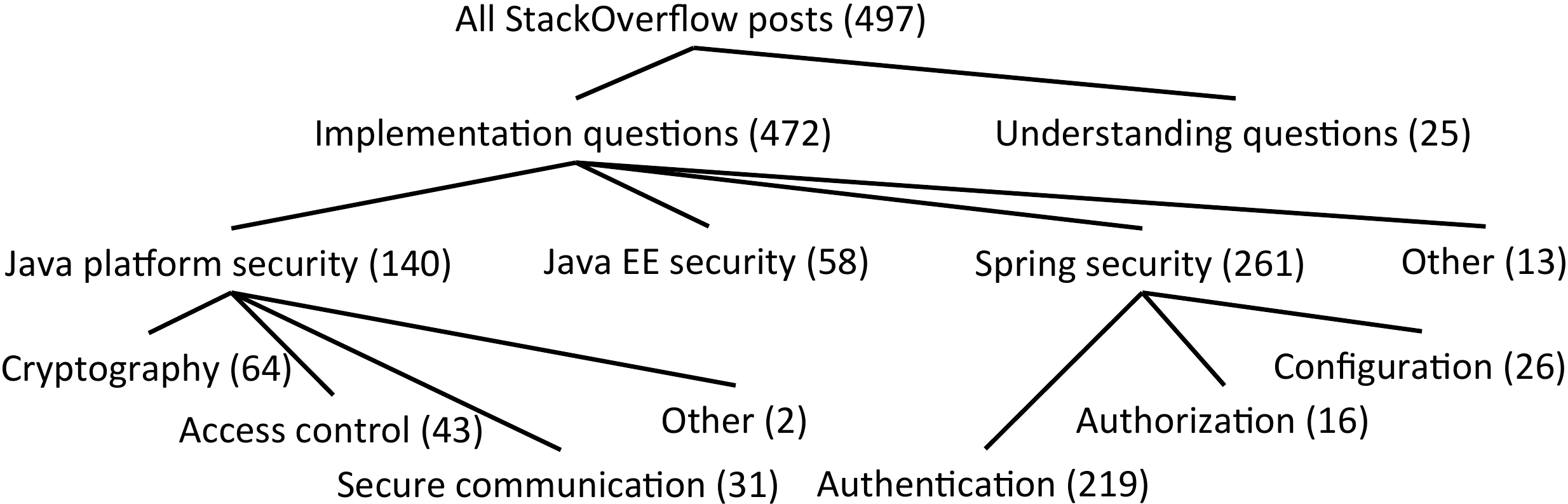}
\caption{Taxonomy of StackOverflow posts}
\label{fig:taxonomy}
\end{figure*}
\vspace{-0.5em}
\paragraph{\textbf{RQ1: What are the common security concerns of developers?}}
We aimed to investigate: (1) what are the popular security features that developers frequently asked about, and (2) how do developers' security concerns shift over the years?
Besides, we also classified posts into three categories based on the number of positive votes and favorite counts their questions received:
\begin{itemize}
\item Neutral: A question does not receive a positive vote or favorite count.
\item Positive: A question receives at least one positive vote but zero favorite count.
\item Favorite: A question receives at least one favorite vote.
\end{itemize}
The post shown in Figure~\ref{fig:post} is classified as ``Favorite'' in this way, because its favorite count is one.
By combining this category with the identified security concerns, we explore developers' sentiment towards questions related to different concerns. It is possible that although some security features are frequently asked, people do not like the questions, probably because the questions are so complicated and project-specific that most developers cannot learn or benefit from them. 

\vspace{-0.5em}
\paragraph{\textbf{RQ2: What are the common programming challenges?}}
For each identified security concern, we further characterized each post with the problem (buggy source code, wrongly implemented configuration files, improperly configured execution environment), the root cause and accepted solution of the problem. Then we clustered posts if they had similar characterizations. For the post in Figure~\ref{fig:post}, we identified the problem as asking for a workaround in SSL verification, because apparently the developer did not realize that the SSL verification should not be bypassed. The recommended solution was to first create a keystore that contains the certificates of all trusted SSL servers, and then use the keystore to create a \codefont{TrustManagerFactory} instance and establish connections.

\vspace{-0.5em}
\paragraph{\textbf{RQ3: What are the common security vulnerabilities?}}
For each post, we also inspected unaccepted answers and the conversational comments between the question asker and other developers, in order to learn about the security context. Based on the recommended security coding practice and the investigated security context, we decided whether the accepted solution was security vulnerable. The post shown in Figure~\ref{fig:post} contains a secure accepted answer, although the question asker originally asked for a vulnerable implementation as an easy fix.

\section{Major Findings}
We present our investigation results for the research questions separately in Section~\ref{sec:trend}-\ref{sec:securityperspectives}.

\subsection{Common Concerns in Security Coding}
\label{sec:trend}
Figure~\ref{fig:taxonomy} presents our classification hierarchy among the \totalcount{} posts. At the highest level of the hierarchy, we created two categories: \textbf{implementation questions} vs.~\textbf{understanding questions}. The majority---\implcount{} questions---were about implementing security functionalities 
or solving program errors. 
Only 25 questions were asked to understand why certain features were designed in certain ways (e.g., how does Java string being immutable increase security~\cite{15274874}).
Since our focus is on secure coding practice, our further classification expands on the \implcount{} implementation-relevant posts. 

At the second level of the classification hierarchy, we clustered posts based on the \emph{major security platforms or frameworks} involved in each post. For instance, \textbf{Java platform security} posts were relevant to the security features of the Java platform and related software libraries extending the features (e.g., BouncyCastle~\cite{bc}). \textbf{Java EE security} posts were related to the security mechanisms of Java EE platform, and \textbf{Spring security} posts were about the Spring security framework~\cite{springsecurity}. The \textbf{Other} category at this level includes posts relevant to other libraries or platforms, such as Shiro~\cite{30463057} and Android~\cite{41278592}. \emph{Unexpectedly, 
Spring security posts (261) counted for 55\% of the implementation questions}.
However, no research exists to explore the misuse of Spring security APIs.
Although there are many application servers developed to implement the Java EE specifications, developers have many fewer questions concerning Java EE security than Spring security. 

At the third level of the classification hierarchy, we further classified posts separately belonging to the two categories: Java platform security and Spring security, because both categories contained many posts. Among the Java platform security posts, in addition to \textbf{cryptography} and \textbf{secure communication}, we identified a third major concern---\textbf{access control}. Among the Spring security posts, we found the majority (219) related to \textbf{authentication}, with the minority discussing \textbf{authorization} and \textbf{configuration}.

\begin{tcolorbox}
	\textbf{Finding 1:}
	\emph{55\%, 30\%, and 12\% of the implementation-relevant posts focused on Spring security, Java platform security, and Java EE security, indicating that developers need more help to secure Java enterprise applications.}
\end{tcolorbox}

Based on the second- and third-level classifications, we identified seven major security concerns: cryptography, access control, secure communication, Java EE security, authentication, authorization, and configuration. 
The first three concerns correspond to Java platform security, while the last three correspond to Spring security. 
To reveal developers' security concern trends over the years, we clustered posts based on the year when each question was asked.

\begin{figure}[!htb]
\centering
\includegraphics[width=8cm]{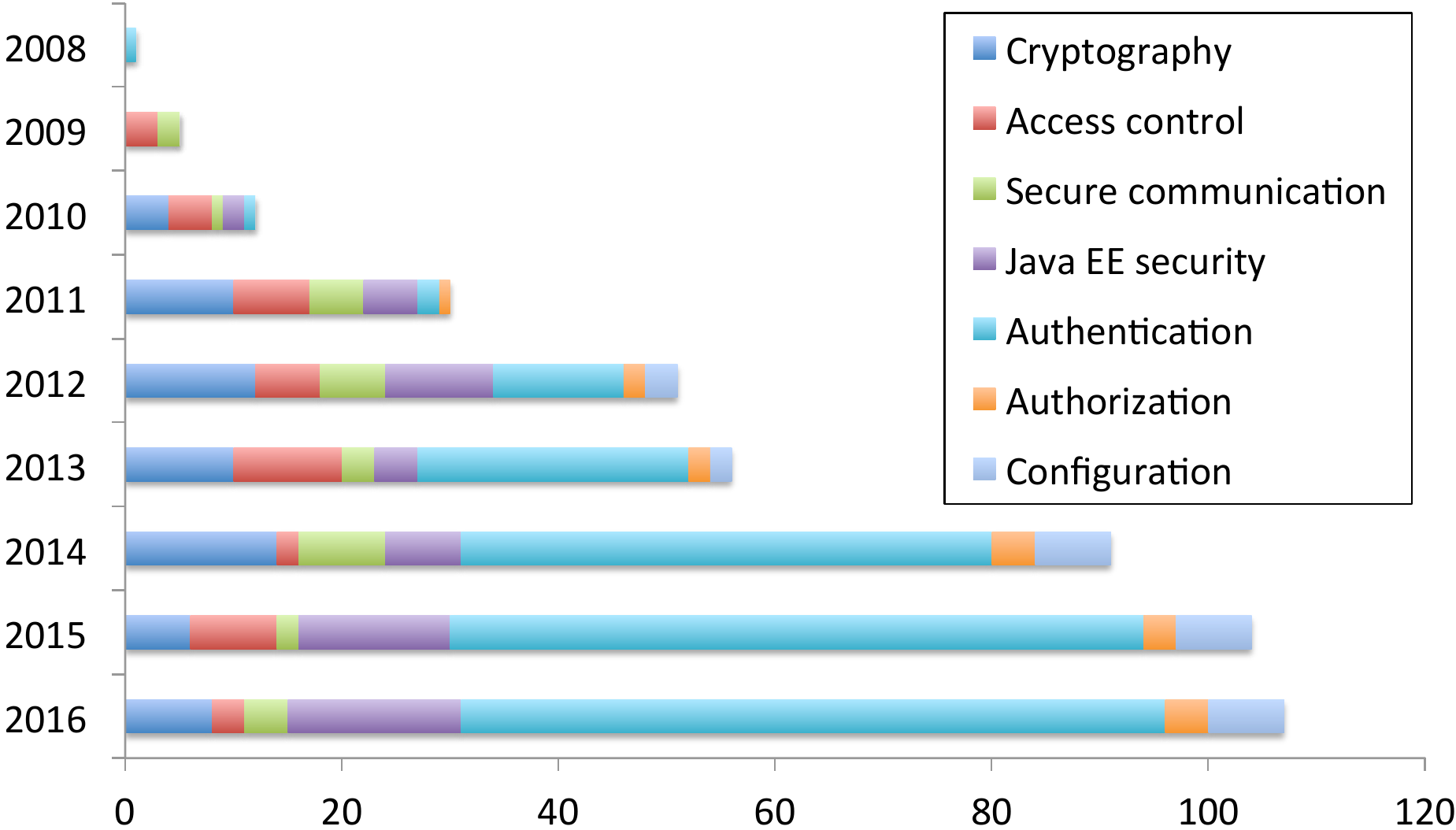}
\vspace{-1em}
\caption{The post distribution during 2008-2016}
\label{fig:year-distribution}
\end{figure}
\vspace{-0.5em}
Figure~\ref{fig:year-distribution} presents the post distribution among 2008-2016. 
The total number of posts increased over the years, indicating that more developers were involved with secure coding and faced problems.
Specifically, 
there was only 1 post created in 2008, but 107 posts were created in 2016. During 2009-2011, most posts were about Java platform security. However, since 2012, the major security concern has shifted to securing Java enterprise applications (including both Java EE security and Spring security). Specifically, Spring security has taken up over 50\% of the posts published every year since 2013.

\begin{figure}[!htb]
\centering
\includegraphics[width=8.5cm]{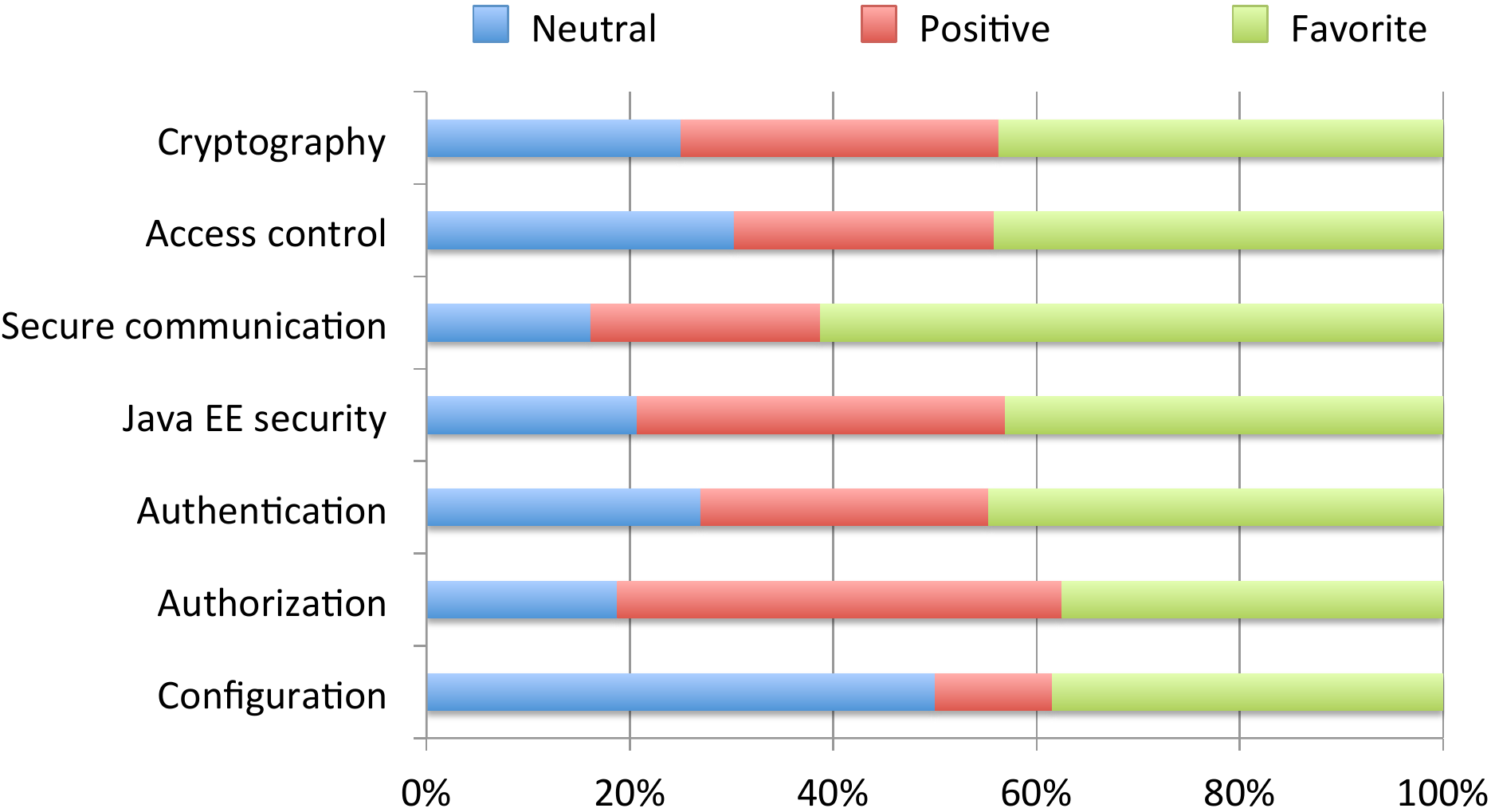}
\vspace{-1em}
\caption{The post distribution among developers' sentiment towards the security features: neutral, positive, and favorite}
\label{fig:attitude-distribution}
\end{figure}

For each security concern, we also clustered posts based on developers' attitudes towards the questions. As mentioned in Section~\ref{sec:method}, we defined three types of sentiment: \emph{neutral}, \emph{positive}, and \emph{favorite}. Figure~\ref{fig:attitude-distribution} shows the post distribution among different developers' attitudes. In this figure, configuration posts received the highest percentage of neutral opinions (50\%). One possible reason is that these posts mainly focused on the problems caused by wrong versions of software libraries and version conflicts between dependent libraries.
Since such problems are usually specific to programmers' software development environments, they are not representative or relevant to many developers' security interests.
In comparison, secure communication posts received the lowest percentage of neutral opinions (16\%), but the highest percentage of favorite (61\%), indicating that the questions were more representative, focusing more on security implementation instead of environment settings.

\begin{tcolorbox}
	\textbf{Finding 2:}
	\emph{Developers' major security concern has shifted from Java platform security to enterprise application security over the years, especially to Spring security. Compared with others, secure communication posts received the highest percentage (61\%) of favorite votes, indicating that the questions are important and representative.}
\end{tcolorbox}

\subsection{Common Programming Challenges}
\label{sec:question}
To understand the common challenges developers faced, 
we further examined the posts of the most popular five major categories: 
authentication (219), cryptography (64), Java EE security (58), access control (43), and secure communication (31).
We identified posts with similar questions and related answers, and further investigated why developers asked these common questions. This section presents our key findings for each category.

\subsubsection{Authentication} 
Most posts were on (1) integrating Spring security with different application servers (e.g., JBoss)~\cite{springjboss} or frameworks (e.g., Spring MVC)~\cite{securityspringmvc}, (2) configuring security in an XML-based~\cite{springxml} or Java-based way~\cite{springjava}, or (3) converting XML-based configurations to Java-based ones~\cite{xmltojava}. Specifically, we observed three challenges. 

\emph{Challenge 1: The way to integrate Spring security with different types of applications varies a lot.} Although Spring security can be leveraged to secure various applications, the usage varies with the application settings~\cite{springvariousapplications}. What is even worse, some Spring security-relevant implementations may exhibit different dynamic behaviors in different applications. For instance, by following a standard tutorial example~\cite{springsample}, a developer defined two custom authentication filters---\codefont{apiAuthenticationFilter} and \codefont{webAuthenticationFilter}---to secure two different sets of URLs of his/her Spring Boot web application as shown below.

\lstset{
numbers=left, 
basicstyle=\small,
escapeinside={(*}{*)},
frame = tb}
\begin{lstlisting}[caption=An exemplar implementation working unexpectedly in Spring Boot applications, label={lst:spring}]
@EnableWebSecurity
public class SecurityConfiguration {
  @Configuration @Order(1)
  public static class ApiConfigurationAdapter 
      extends WebSecurityConfigurerAdapter {
    @Bean 
    public GenericFilterBean 
        apiAuthenticationFilter() {...}      
    @Override 
    protected void configure(HttpSecurity http) 
        throws Exception {
      http.antMatcher("/api/**")
        .addFilterAfter(apiAuthenticationFilter()...)
        .sessionManagement()...;  } }
  @Configuration @Order(2)
  public static class WebSecurityConfiguration 
      extends WebSecurityConfigurerAdapter {
    @Bean
    public GenericFilterBean 
        webAuthenticationFilter() {...}
    @Override
    protected void configure(HttpSecurity http) 
        throws Exception {
      http.antMatcher("/")
          .addFilterAfter(webAuthenticationFilter()...)
          .authorizeRequests()...; } } }
\end{lstlisting}
In Listing~\ref{lst:spring}, lines 3-14 correspond to \codefont{ApiConfigurationAdapter}, a security configuration class that specifies \codefont{apiAuthenticationFilter} to authenticate URLs matching the pattern ``/api/**''. Lines 15-26 correspond to \codefont{WebSecurityConfiguration}, which configures \codefont{webAuthentication\-Filter} to authenticate the other URLs. Ideally, only one filter is invoked given one URL. However, both filters were invoked in reality. The root cause is that each filter is a bean (annotated with @Bean on lines 6 and 18). Spring Boot detects the filters and adds them to a regular filter chain, while Spring security also adds them to its own filter chain. Consequently, both filters are registered twice and can be invoked twice. To solve the problem, developers need to enforce each bean to be registered \emph{only once} by adding specialized code.

\emph{Challenge 2: The two ways of security configurations (Java-based and XML-based) are hard to implement correctly.} Take the Java-based configuration for example. There are lots of annotations and APIs of classes, methods, and fields available to specify different configuration options. Particularly, \codefont{HttpSecurity} has 10 methods, each of which can be invoked on an \codefont{HttpSecurity} instance and then produces another \codefont{HttpSecurity} object. If developers are not careful about the invocation order between these methods, they can get errors~\cite{securitymethodorder}. As shown in Listing~\ref{lst:spring}, the method \codefont{antMatcher("/api/**'')} must be invoked {\em before} \codefont{addFilterAfter(...)} (lines 12-13), so that the filter is only applied to URLs matching the pattern ``/api/**''. Unfortunately, such implicit constraints are not documented in the API specifications~\cite{springsecuritydoc}.

\emph{Challenge 3: Converting from XML-based to Java-based configurations is tedious and error-prone.} 
The semantic conflicts between annotations, deployment descriptors, and code implementations are always hard to locate and resolve. Such problems become more serious when developers configure security in a hybrid way of Java-based and XML-based. Since Spring security 3.2, developers are supported to configure Spring-security in a pure Java-based approach, and there is documentation describing how to migrate from XML-based to Java-based configurations~\cite{springsecurityreference32}. However, manually applying every migration rule to convert every configuration file is tedious and error-prone.


\begin{tcolorbox}
	\textbf{Finding 3:}
	\emph{Spring security authentication posts were mainly about configuring security for different applications or in different ways (Java-based or XML-based), and converting between these ways. The challenges were due to incomplete documentation and missing tool support for error checking, solution recommendation, and configuration generation.
	}
\end{tcolorbox}

\subsubsection{Cryptography} 
Many posts were about key generation and usage. For instance, some posts discussed how to create a key from scratch~\cite{keyfromscratch}, or how to generate or retrieve a key from a random number~\cite{keyfromnumber}, a byte array~\cite{keyfrombytes}, a string~\cite{keyfromstring}, a certificate~\cite{keyfromcertificate}, BigIntegers~\cite{keyfrombigintegers}, a keystore~\cite{keyfromkeystore}, or a file~\cite{keyfromfile}. Some other posts focused on how to compare keys~\cite{keyforcomparison}, print key information~\cite{keyforprint}, or initialize a cipher for encryption and decryption~\cite{keyforcipher}. Specifically, we observed three common challenges of correctly using the cryptography APIs. 

\emph{Challenge 1: The error messages did not provide sufficient useful hints about fixes.} We found five posts concentrated on the same problem: ``get InvalidKeyException: Illegal key size'', while the solutions were almost identical: (1) download the ``Java Cryptography Extension (JCE) Unlimited Strength Jurisdiction Policy Files'', ``local\_policy.jar'', and ``US\_export\_policy.jar''; and (2) place the policy files in proper folders~\cite{keycommonerror}. Developers got the same exception because of missing either of the two steps. Providing a checklist of these necessary steps in the error message could help developers quickly resolve the problem. However, the existing error messages did not provide any constructive suggestion.


\emph{Challenge 2: It is difficult to implement security with multiple programming languages.}
Three posts were about encrypting data with one language (e.g. PHP or Python) and decrypting data with another language (e.g., Java). Such cross-language data encryption \& decryption is challenging, because the format of the generated data by one language requires special handling of another language. Listing~\ref{lst:keypair} is an example to generate an RSA key pair and encrypt data in PHP, and to decrypt data in Java~\cite{commonencryptionerror}.
\lstset{
numbers=left, 
basicstyle=\small,
escapeinside={(*}{*)},
frame = tb}

\begin{lstlisting}[caption=Encryption in PHP and decryption in Java, label={lst:keypair}]
(*\bf // ***** keypair.php ******)
if(file_exists('private.key')) {
    echo file_get_contents('private.key'); }
else {
    include('Crypt/RSA.php');
    $rsa = new Crypt_RSA();
    $res = $rsa->createKey();
    $privateKey = $res['privatekey'];
    $publicKey  = $res['publickey'];
    file_put_contents('public.key',$publicKey);
    file_put_contents('private.key',$privateKey); }
(*\bf // ***** encrypt.php ******)
include('Crypt/RSA.php');
$rsa = new Crypt_RSA();
$rsa->setEncryptionMode(CRYPT_RSA_ENCRYPTION_OAEP);
$rsa->loadKey(file_get_contents('public.key'));
(*\bf // ***** MainClass.java ******)
BASE64Decoder decoder=new BASE64Decoder();
String b64PrivateKey=getContents(
  "http://localhost/api/keypair.php").trim();
byte[] decodedKey=decoder.decodeBuffer(b64PrivateKey);
BufferedReader br=new BufferedReader(
  new StringReader(new String(decodedKey)));
PEMReader pr=new PEMReader(br);
KeyPair kp=(KeyPair)pr.readObject();
pr.close();
PrivateKey privateKey=kp.getPrivate();
Cipher cipher=Cipher.getInstance(
  "RSA/None/OAEPWithSHA1AndMGF1Padding","BC");
cipher.init(Cipher.DECRYPT_MODE, privateKey);
byte[] plaintext=cipher.doFinal(cipher);
\end{lstlisting}
In this example, when a key pair is generated in PHP (lines 2-11),
the public key is easy to retrieve in PHP (lines 13-16). 
However, retrieving the private key in Java is more complicated (lines 18-30).
After reading in the private key string (lines 19-20), the Java implementation first uses \codefont{Base64Decoder} to decode the string into a byte array (line 21), which corresponds to an OpenSSL PEM encoded stream (line 22-23). 
Because OpenSSL PEM is not a standard data format, the Java code further uses a PEMReader to convert the stream to a PrivateKey instance (lines 24-27) before using the key to initialize a cipher (lines 28-30).
Existing documentation seldom describes how the security data format (e.g., key) defined in one language corresponds to that of another language. Unless developers are experts in both languages, it is hard for them to figure out the security data processing across languages.

\emph{Challenge 3: Implicit constraints on API usage cause confusion.}
Two posts were about getting ``InvalidKeySpecException: algid parse error, not a sequence'', when obtaining a private key from a file~\cite{pkcs8}. The problem is that the key should be in PKCS\#8 format when used to create a \codefont{PKCS8EncodedKeySpec} instance, as shown below:

\lstset{
frame = tb}

\begin{lstlisting}[caption=Consistency between the key format and keyspec]
//privKey should be in PKCS#8 format
byte[] privKey=...;
PKCS8EncodedKeySpec keySpec=
  new PKCS8EncodedKeySpec(privKey); 
\end{lstlisting}
The tricky part here is that a private key retrieved from a file always has the data type \codefont{byte[]} even if it is not in PKCS\#8 format. If developers invoke the API \codefont{PKCS8EncodedKeySpec(...)} with a non-PKCS\#8 formatted key, they will be stuck with the clueless exception. Three solutions were suggested to get a PKCS\#8 format key: (1) to implement code to convert the byte array, (2) to use an openssl command to convert the file format, or (3) to use the \codefont{PEMReader} class of BouncyCastle to generate a key from the file. Such implicit constraints between an API and its input format are delicate. 



\begin{tcolorbox}
	\textbf{Finding 4:}
	\emph{The cryptography posts were majorly about key generation and usage. Developers asked these questions mainly due to clueless error messages, cross-language data handling, and implicit API usage constraints.}
\end{tcolorbox}

\subsubsection{Java EE security}\label{sec:javaeesecurity} 33 of the 58 posts were on authentication and authorization. However, the APIs of these two security features were defined differently on different application servers (e.g., WildFly and Glassfish), and developers might use these servers in combination with diverse third-party libraries~\cite{javaeecombination}. As a result, the posts seldom shared common solutions or code implementation.  

One common challenge we identified is the usage of declarative security and programmatic security. When developers misunderstood annotations, they could use incorrect annotations that conflict with other annotations~\cite{conflictannotations}, deployment descriptors~\cite{conflictannotationdescriptor}, code implementation~\cite{conflictannotationcode}, or file paths~\cite{conflictannotationpath}. Nevertheless, existing error reporting systems only throw exceptions. Unfortunately, there is no tool support that prevents developers from configuring such conflicting settings, or assists developers with diagnosing conflicting usage of annotations and deployment descriptors.
\begin{tcolorbox}
	\textbf{Finding 5:}
	\emph{Java EE security posts were mainly about authentication and authorization. 
One challenge is the complex usage of declarative security and programmatic security, and any complicated interaction between the two. 	
	}
\end{tcolorbox}

\subsubsection{Access Control} The 43 posts mainly discussed how to \emph{restrict} or \emph{relax} the access permission(s) of a software application for certain resource(s). 

Specifically, 21 questions asked about restricting untrusted code from accessing certain packages~\cite{limitpackageaccess}, classes~\cite{limitclassaccess}, or class members (i.e., methods and fields)~\cite{limitmemberaccess}. 
Two alternative solutions were commonly suggested for these questions: (1) to override the \codefont{checkXXX()} methods of \codefont{SecurityManager} to disallow invalid accesses, or (2) to define a custom policy file to grant limited permissions. Another nine posts were on how to allow applets to perform privileged operations~\cite{allowapplet}, because applets are executed in a security sandbox by default and can only perform a set of safe operations. One commonly recommended solution was to digitally sign the applet. Although it seems that there exist common solutions to the most frequently asked questions, the access control implementation is not always intuitive.
We identified two common challenges of correctly implementing access control.

\emph{Challenge 1: The effect of access control varies with the program context.}
We identified two typical scenarios from multiple similar posts. 
First, the RMI tutorial~\cite{rmitutorial} suggested that a security manager is needed \emph{only} when RMI code downloads code from a remote machine. Including a \codefont{SecurityManager} instance in the RMI program which does not download any code can cause an AccessControlException~\cite{rmism}. 
Second, although a signed applet is allowed to perform sensitive operations, it loses its privileges when invoked from Javascript~\cite{appletjs}. As a result, the invocation to the signed applet should be wrapped with an invocation of \codefont{AccessController. doPrivileged(...)}.

\emph{Challenge 2: The effect of access control varies with the execution environment.}
\codefont{SecurityManager} can disallow illegal accesses via reflection only when the program is executed in a controlled environment (i.e., on a trusted server)~\cite{reflectionenvironment}. Nevertheless, if the program is executed in an uncontrolled environment (e.g. on an untrusted client machine) and hackers can control how to run the program or manipulate the jar file, the security mechanisms become voided.

\begin{tcolorbox}
	\textbf{Finding 6:}
	\emph{The access control posts were mainly about \codefont{SecurityManager}, \codefont{AccessController}, and the policy file. Configuring and customizing access control policies are challenging.}
\end{tcolorbox}

\subsubsection{Secure Communication} Among the 31 examined posts, 22 posts were about SSL/TLS-related issues, discussing how to create~\cite{certcreate}, install~\cite{certinstall}, 
find~\cite{sslcertificate}, 
or validate an SSL certificate~\cite{certuse}, how to establish a secure connection~\cite{establishssl}, and how to use SSL together with other libraries, such as JNDI~\cite{ssljndi} and PowerMock~\cite{sslpowermock}.

In particular, six posts focused on solving the problem of unable to find a valid server certificate to establish an SSL connection with a server~\cite{sslcertificate}. 
Instead of suggesting a way to install the required certificates, two accepted answers suggested a highly insecure workaround to disable the SSL verification process, so that any incoming certificate can pass the validation~\cite{sslworkaround}. Although such workarounds can effectively remove the error, they essentially fail the requirement to secure communication with SSL. In Section~\ref{sec:securityperspectives}, we will further explain the security vulnerability due to such workarounds. Probably developers tended to accept the vulnerable answers because they felt it challenging to implement the whole process of creating, installing, finding, and validating an SSL certificate.

\begin{tcolorbox}
	\textbf{Finding 7:}
	\emph{Security communication posts mainly discussed the process of establishing SSL/TLS connections. This process contains so many steps that developers were tempted to accept a broken solution to simply bypass the security check.}
\end{tcolorbox}

\subsection{Common Problems from Security Perspectives}
\label{sec:securityperspectives}

Among the five categories listed in Section~\ref{sec:question}, we identified security vulnerabilities in the accepted answers of three frequently discussed topics: 
Spring security's \codefont{csrf()}, SSL/TLS, and password hashing.

\subsubsection{Spring security's \codefont{csrf()}}
\textbf{Cross-site request forgery (CSRF)} 
is a serious attack that tricks a web browser into executing an unwanted action (e.g., transfer money to another account) in a web application (e.g., a bank website) for which a user is authenticated~\cite{zellercross}. The root cause is that attackers created forged requests that appear to be legitimate requests, and somehow mixed them with the legitimate ones. Since the application cannot distinguish between the two types of requests, it normally responds to the forged requests, performing undesired operations. 

By default, Spring security provides CSRF protection by defining a function \codefont{csrf()} and implicitly enabling the function invocation. Correspondingly, developers should include the CSRF token in all PATCH, POST, PUT, and DELETE methods to leverage the protection~\cite{csrftoken}. However, among the 12 examined posts that were relevant to \codefont{csrf()}, 5 posts discussed program failures, while all the accepted answers suggested an insecure solution: disabling the CSRF protection by invoking \codefont{http.csrf().disable()}. In one instance, after accepting the vulnerable solution, an asker commented as ``\emph{Adding csrf().disable() solved the issue!!! I have no idea why it was enabled by default}''~\cite{csrfdisable}. Unfortunately, the developer happily disabled the security protection without realizing that such workaround would expose the resulting system to CSRF security exploits. 
 
\begin{tcolorbox}
	\textbf{Finding 8:}
	\emph{In 5 of the 12 \codefont{csrf()}-relevant posts, developers took the suggestion to irresponsibly disable the default CSRF protection. Developers are unaware of the security consequences of their insecure coding.}
\end{tcolorbox}

\subsubsection{SSL/TLS}
We examined 11 posts discussing the usage of SSL/TLS, and observed two important security issues.

\begin{figure}[!htb]
\centering
\includegraphics[width=6cm]{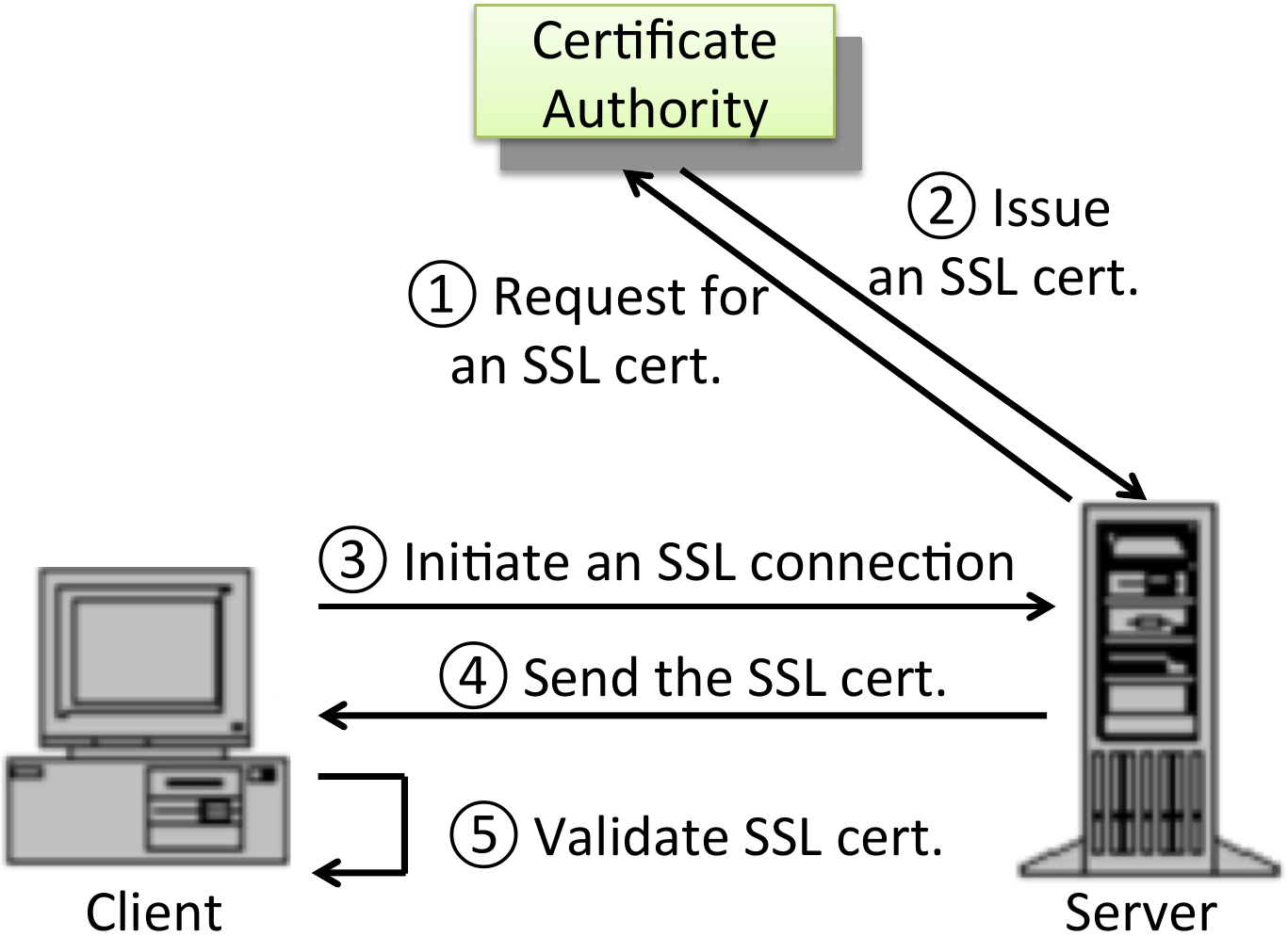}
\vspace{-1em}
\caption{Simplified overview of creating an SSL connection}
\label{fig:ssloverview}
\end{figure}
\vspace{-1em}

\emph{Problem 1: Developers commonly trusted all SSL certificates and allowed all hostnames in order to quickly build a prototype in the development environment.} SSL is the standard security technology for establishing an encrypted connection between a web server and a browser. 
Figure~\ref{fig:ssloverview} overviews the major steps of establishing an SSL connection~\cite{rfs6101}. To activate SSL on the server, developers need to provide all identity information of the website (e.g., the host name) to a Certification Authority (CA) and request for an SSL certificate (Step \textcircled{1}). After validating the website's information, CA issues a digitally signed SSL certificate (Step \textcircled{2}). When a client or browser attempts to connect to the website ((Step \textcircled{3}), the server sends over its certificate (Step \textcircled{4}). The client then conducts several checks, including (1) whether the certificate is issued by a CA the browser trusts, and (2) whether the request hostname matches the hostname associated with the certificate (Step \textcircled{5}).
If all these checks are passed, the SSL connection can be established successfully.

Although ideally, developers should only enable SSL after obtaining a certificate from CA, in reality, they usually implement and test the certificate verification code before obtaining the certificate. Therefore, a well-accepted recommended solution without CA-signed certificates is to create a self-signed certificate and use the certificate to drive the implementation of SSL certificate verification~\cite{certcreate}. However, 9 of the 11 examined posts accepted an insecure solution to bypass security checks by trusting {\em all} certificates and/or allowing {\em all} hostnames, as demonstrated by Listing~\ref{lst:sslbroken}.  

\lstset{
basicstyle=\small,
breaklines=true, 
frame = tb}
\begin{lstlisting}[caption=A typical implementation to disable SSL certificate validation~\cite{ssltypical}, label={lst:sslbroken}]
(*\bfseries // Create a trust manager that does not validate certificate chains*)
TrustManager[] trustAllCerts = new TrustManager[]{
  new X509TrustManager() {
    public java.security.cert.X509Certificate[] getAcceptedIssuers() {return null;}
    public void checkClientTrusted(...) {}
    public void checkServerTrusted(...) {} }};
(*\bfseries // Install the all-trusting trust manager*)
try {
  SSLContext sc = SSLContext.getInstance("SSL");
  sc.init(null, trustAllCerts, new java.security.SecureRandom());
  HttpsURLConnection.setDefaultSSLSocketFactory(sc.getSocketFactory());
} catch (Exception e) {}
(*\bfseries// Access an https URL without having the certificate in the*)
(*\bfseries// truststore*)
try {
  URL url=new URL("https://hostname/index.html");
} catch (MalformedURLException e) {}
\end{lstlisting} 

Disabling the SSL certificate validation process in a client can thoroughly invalidate the secure communication protocol, leaving clients susceptible to the \textbf{man-in-the-middle (MITM) attack}~\cite{Georgiev:2012}. Namely, by secretly relaying and possibly altering the communication (e.g., through DNS poisoning) between the client and server, attackers can trick the SSL-client to instead connect to an attacker-controlled server~\cite{Georgiev:2012}. Although the insecure coding practice was shown to induce the MITM attack in 2012~\cite{Georgiev:2012}, there are three examined posts created after 2012 still discussing the bad practice. This indicates a significant gap between security theory and coding practices. Some developers justified their checking-bypass logic by saying ``\emph{Because I needed a quick solution for debugging purposes only. I would not use this in production due to the security concerns \ldots}''~\cite{ssldebate}. However, as pointed by another user of StackOverflow~\cite{ssldebate} and demonstrated by prior research~\cite{Georgiev:2012,Fahl:2012}, \textbf{\emph{a lot of these implementations find their way into production, and have yielded radically insecure systems as a result}}. 

\emph{Problem 2: Developers were unaware of the best usage of SSL/TLS.}
TLS is SSL's successor. It is so different from SSL that the two protocols do not interoperate. To maintain the backwards compatibility with SSL 3.0 and interoperate with systems supporting SSL, most SSL/TLS implementations allow for protocol version negotiation: if a client and a server cannot connect via TLS, they will fall back to using the older protocol SSL 3.0.
In 2014, Möller et al.~reported the \textbf{POODLE attack} which exploits the SSL 3.0 fallback~\cite{moller2014poodle}. Specifically, there is a design vulnerability in the way SSL 3.0 handles block cipher mode padding, which can be exploited by attackers to decrypt encrypted messages. With the POODLE attack, a hacker can intentionally trigger a TLS connection failure and force usage of SSL 3.0, allowing decryption of encrypted messages. 

Ever since 2014, researchers have recommended developers to disable SSL 3.0 support and configure systems to present the SSL 3.0 fallback. The US government (NIST) mandates to ceasing usage of SSL in the protection of Federal information~\cite{nist}. In reality, nevertheless, none of the 11 posts mentioned the POODLE attack. Both of the two examined posts created in 2016 were about SSL usage.  

\begin{tcolorbox}
	\textbf{Finding 9:}
	\emph{9 of 11 SSL/TLS-relevant posts discussed insecure code to bypass security checks. We observed two important security threats: (1) StackOverflow contains a lot of obsolete and insecure coding practices; and (2) secure programmers are unaware of the state-of-the-art security knowledge.}
\end{tcolorbox}

\subsubsection{Password Hashing}
We found six posts related to hashing passwords with MD5 or SHA-1 to store the user credentials in a database. 
However, these hashing functions were found insecure~\cite{md5wang04,sha1stevens17}.
They are vulnerable to offline \textbf{dictionary attacks}~\cite{dictionaryattack}. After obtaining a password hash $H$ from a compromised database, a hacker can use brute-force methods to enumerate a list of password guesses, until finding the password $P$ whose hash value is $H$. By impersonating a valid user to login a server, the attacker can conduct malicious behaviors.
Researchers recommended \textbf{key-stretching algorithms} (e.g., PBKDF2, bcrypt, and scrypt) as the best practice for secure password hashing, because these algorithms are specially crafted to slow down the hash computation by orders of magnitude~\cite{owaspcheatsheet17, gackenheimer2013implementing, boonkrong2012security}, which substantially increases the difficulty of dictionary attacks.

Unfortunately, only three of the six posts (50\%) mentioned the best practice in the accepted answers, indicating that many posts on secure hashing suggested insecure hash functions.
We found one post which asked about using MD5 hashing in Android~\cite{md5}. Within the comment conversation between developers, although people recommended to avoid MD5, the asker kept justifying his/her choice of MD5. The asker even shared a completely wrong understanding of secure hashing: ``\emph{The security of hash algorithms \textbf{really} is MD5 (strongest) > SHA-1 > SHA-256 > SHA-512 (weakest)}'', although the opposite is true, which is MD5 < SHA-1 < SHA-256 < SHA-512.
Among these posts, some developers misunderstood security APIs and ignored the security consequences of their security API choices. Such StackOverflow posts can have profound negative impact, because they convey the wrong information and may mislead people. 

\begin{tcolorbox}
	\textbf{Finding 10:}
	\emph{Three of six hashing-relevant posts accepted vulnerable solutions as correct answers, indicating that developers were unaware of the best practice of secure programming. Their wrong knowledge or practice can propagate among StackOverflow users and negatively influence people.}
\end{tcolorbox}

\section{Related Work}
This section describes related work on analyzing, detecting, and preventing security vulnerabilities due to Java library misuse.

\subsection{Analyzing Security Vulnerabilities} Prior studies showed that the API misuse of cryptography, SSL, and Java reflection caused many security vulnerabilities~\cite{LongSoftwareVulnerabilities2005,Lazar:2014,veracode, Yang2016}. For instance, Long identified several Java features whose misuse or improper implementation can compromise security~\cite{LongSoftwareVulnerabilities2005}. One feature he identified is the Java reflection API which enables fields that are not normally accessible to be accessed and thus can cause potential vulnerabilities.
Lazar et al.~manually examined 269 published cryptographic vulnerabilities in the CVE database, and observed 83\% of them were caused by the misuse of cryptographic libraries, including low encryption strength, insufficient randomness, and inadequate checks~\cite{Lazar:2014}. Veracode reported that 39\% of all applications used broken or risky cryptographic algorithms~\cite{veracode}. The study by Yang et al.~\cite{Yang2016} is most relevant to our research. They used an advanced topic model approach, Latent Dirichlet Allocation (LDA) tuned using Genetic Algorithm (GA), to cluster security-related StackOverflow questions based on the text. They identified frequently mentioned security topics like ``Password'' and ``Hash'' for categorization purposes. In comparison, we investigate the programming challenges and security vulnerabilities among the topics. Our SE and security findings have more technical depth.

\subsection{Detecting Security Vulnerabilities}
Approaches were built to detect security vulnerabilities caused by API misuse~\cite{Fahl:2012,Georgiev:2012,Egele:2013,Li2014,Onwuzurike:2015,He2015,Chatzikonstantinou:2016,fischer:2017}. For instance, Egele et al.~implemented a static checker for six well defined Android cryptographic API usage rules, such as ``Do not use ECB mode for encryption'', and analyzed 11,748 Android applications for any rule violation~\cite{Egele:2013}. They found 88\% of the applications violating at least one checked rule.
Fischer et al.~extracted Android security-related code snippets from StackOverflow, and manually labeled a subset of the data as ``secure'' or ``insecure''~\cite{fischer:2017}. The labeled data allowed them to train a classifier and efficiently judge whether a code snippet is secure or not for the whole data set. Next, they searched for code clones of the snippets in 1.3 million Android apps, and found many clones of the insecure code.
Fahl et al.~\cite{Fahl:2012} and Georgiev et al.~\cite{Georgiev:2012} separately implemented an attack model: man-in-the-middle attack, and detected vulnerable Android applications and popular software libraries which misused SSL APIs. Both research groups observed that developers disabled certification validation for testing with self-signed and/or trusted certificates.
He et al.~developed SSLINT, an automatic static analysis tool, to identify the misuse of TLS/SSL APIs in client-side applications~\cite{He2015}. 


Compared with prior research, our study has two new contributions. 
First, \emph{our scope is broader.} We report new challenges on secure coding practices, such as complex security configurations in Spring security, poor error messages, and multilingual programs.
Second, \emph{our investigation on the online forum provides a new social and community perspective about secure coding}. The unique insights cannot be discovered through analyzing code.  



\subsection{Preventing Security Vulnerabilities}
Various techniques were proposed to prevent developers from implementing vulnerable code and misusing APIs~\cite{MettlerWagnerClose10,keyczar,Mitchell:1997,Datta2007,Smith2008,Erkok:2009}. For example, Mettler et al.~designed Joe-E---a security-oriented subset of Java---to support secure software development by removing any encapsulation-breaking features from Java (e.g., reflection), and by enforcing the least privilege principle (i.e., by default, each Joe-E object has no privilege to access system resources, unless another entity passes it a reference to a system resource object)~\cite{MettlerWagnerClose10}. Keyczar is a library designed to simplify the usage of cryptography, and thus to prevent API misuse~\cite{keyczar}. Below shows how Keyczar APIs are used to decrypt data:
\begin{lstlisting}[caption=Simple decryption with Keyczar APIs]
Crypter crypter=new Crypter("/rsakeys");
String plaintext=crypter.decrypt(ciphertext);
\end{lstlisting}
Compared with the decryption code shown in Listing~\ref{lst:keypair} (lines 18-31), this implementation is much simpler and more intuitive. All details about data format conversion and cipher initialization are hidden, while a default strong block cipher is used to properly decrypt data.

Some approaches were developed to apply formal verification techniques and analyze the security properties of cryptographic protocol specifications~\cite{Mitchell:1997,Datta2007} and cryptographic API implementations~\cite{Smith2008,Erkok:2009}. For instance, Protocol Composition Logic (PCL) is a logic for proving security properties, like network protocols that use public and symmetric key cryptography~\cite{Datta2007}. The logic is designed around a process calculus with actions for possible protocol steps including generating new random numbers, sending and receiving messages, and performing decryption and digital signature verification actions. The proof system consists of axioms about individual protocol actions and inference rules that yield assertions about protocols composed of multiple steps.

\section{Our Recommendations}

By analyzing the StackOverflow posts relevant to Java security from both software engineering and security perspectives, we observed the gap between the intended usage of APIs and the actual problematic API usage by developers, sensed developers' frustration when they spent tremendous time figuring out the correct usage of APIs (e.g., two weeks as mentioned in~\cite{securityspringmvc}), and observed terrible security consequences of library misuse. Below are our recommendations based on the analysis.

\paragraph{\textbf{For Security Developers}}
Conduct security testing to check whether the implemented features work as expected. 
Do not disable security checks (e.g., SSL certificate validation) to implement a temporary workaround in the testing or development environment.
Be cautious when following the StackOverflow accepted answers to implement secure code, because these solutions may be unsafe and outdated. For administrators of StackOverflow, we recommend them to carefully handle the posts that suggest vulnerable code, because these posts can play an influential negative role when educating security programmers.

\paragraph{\textbf{For Library Designers}} 
Remove or deprecate the APIs whose security protection is broken (e.g., MD5). 
Design clean and helpful error reporting interfaces which show not only the error, but also the possible root causes and solutions.
Design simplified APIs with strong security defenses implemented by default.

\paragraph{\textbf{For Tool Builders}}
Develop automatic tools to diagnose security errors, locate buggy code, and suggest security patches or solutions.
Build vulnerability prevention techniques that compare peer applications that use the same set of APIs to infer and warn potential misuses.
Explore approaches that check and enforce the semantic consistency between security-relevant annotations, code, and configurations. 
Build new approaches to transform between the implementations of declarative security and programmatic security.

\section{Threats to Validity}
This study is mainly based on our manual inspection of Java security-relevant posts, so the observations may be subject to human bias. To alleviate the problem, the first author of the paper carefully inspected all posts relevant to implementation questions multiple times, while the second author also examined the posts related to security vulnerabilities (mentioned in Section~\ref{sec:securityperspectives}) multiple times. 

To remove posts without code snippets, we defined a filter to search for keywords ``public'' and ``class''. If a post does not contain both words, the filter automatically removes the post from our data set. This filter may incorrectly remove some relevant posts that contain code. In future, we will improve our crawling technique to keep the <code> tags around code snippets in the raw data, and then rely on these tags to filter posts more precisely.

We conservatively mentioned posts whose accepted answers will cause security vulnerabilities, although there might be more accepted answers that suffer from known security attacks. Due to the limited available program and environment information in each post, and our limited knowledge about frameworks and potential security attacks, we decided not to mention the suspicious posts whose accepted answers might lead to security vulnerabilities.

\section{conclusion}

Our work aimed at assessing the current secure coding practices, and identifying potential gaps between theory and practice and between specification and implementation. Our analysis of hundreds of posts on the popular developer forum (StackOverflow) revealed a worrisome reality in the software development industry. 

\begin{itemize}

\item

 A substantial number of developers do not appear to understand the security implications of coding options, showing a lack of cybersecurity training. This situation creates frustration in developers, who sometimes end up choosing completely insecure-but-easy fixes. Examples of such easy fixes include using obsolete cryptographic hash functions, disabling cross-site request forgery protection, trusting all certificates in HTTPS verification, or using obsolete communication protocols.  These poor coding practices, if used in production code, will seriously compromise the security of software products. 

\item

We provided substantial empirical evidences showing that APIs in Spring security (designed for enterprise security applications) are overly complicated and poorly documented, and error reports from runtime systems cause confusion. In addition, multi-language support for securing data is rather weak. The multi-language situation is common in security applications, as oftentimes the data is encrypted in one programming language (e.g., Python) and decrypted in another (e.g., Java). These issues can seriously hinder developers' productivity, resulting in great frustration and confusion. 

\item

Interestingly, we found that the social dynamics among askers and responders can impact people's security choices. Some influential posts are not secure. For example, we observed that in some cases an insecure suggestion by a user with a high reputation score was selected as the accepted answer, as opposed to the correct fix by a user with a lower reputation score~\cite{reputation}. Sometimes insecure answers have many positive StackOverflow votes (as the quick fixes indeed make  error messages go away)~\cite{ssltypical}, which is quite misleading. 

\end{itemize}

We described several possible solutions to improve secure coding practices in the paper. However, efforts (e.g., workforce retraining) to correct these alarming security issues  may take a while to take effect. Our future work is on building automatic or semi-automatic security bug detection and fixing tools. 

\bibliographystyle{ACM-Reference-Format}
\bibliography{sigproc} 


\begin{thebibliography}{00}


\ifx \showCODEN    \undefined \def \showCODEN     #1{\unskip}     \fi
\ifx \showDOI      \undefined \def \showDOI       #1{#1}\fi
\ifx \showISBNx    \undefined \def \showISBNx     #1{\unskip}     \fi
\ifx \showISBNxiii \undefined \def \showISBNxiii  #1{\unskip}     \fi
\ifx \showISSN     \undefined \def \showISSN      #1{\unskip}     \fi
\ifx \showLCCN     \undefined \def \showLCCN      #1{\unskip}     \fi
\ifx \shownote     \undefined \def \shownote      #1{#1}          \fi
\ifx \showarticletitle \undefined \def \showarticletitle #1{#1}   \fi
\ifx \showURL      \undefined \def \showURL       {\relax}        \fi
\providecommand\bibfield[2]{#2}
\providecommand\bibinfo[2]{#2}
\providecommand\natexlab[1]{#1}
\providecommand\showeprint[2][]{arXiv:#2}

\bibitem[\protect\citeauthoryear{??}{key}{[n. d.]a}]%
        {keycommonerror}
 \bibinfo{year}{[n. d.]}\natexlab{a}.
\newblock \bibinfo{title}{{AES-256 implementation in GAE}}.
\newblock
  \bibinfo{howpublished}{\url{https://stackoverflow.com/questions/12833826/aes-256-implementation-in-gae}}.
    (\bibinfo{year}{[n. d.]}).
\newblock


\bibitem[\protect\citeauthoryear{??}{asr}{[n. d.]}]%
        {asref}
 \bibinfo{year}{[n. d.]}\natexlab{}.
\newblock \bibinfo{title}{Apache Shiro Documentation}.
\newblock
  \bibinfo{howpublished}{\url{https://shiro.apache.org/documentation.html}}.
  (\bibinfo{year}{[n. d.]}).
\newblock


\bibitem[\protect\citeauthoryear{??}{web}{[n. d.]a}]%
        {weblogic}
 \bibinfo{year}{[n. d.]}\natexlab{a}.
\newblock \bibinfo{title}{{Application Server - Oracle WebLogic Server}}.
\newblock
  \bibinfo{howpublished}{\url{https://www.oracle.com/middleware/weblogic/index.html}}.
    (\bibinfo{year}{[n. d.]}).
\newblock


\bibitem[\protect\citeauthoryear{??}{304}{[n. d.]}]%
        {30463057}
 \bibinfo{year}{[n. d.]}\natexlab{}.
\newblock \bibinfo{title}{{authc filter is not calling MyRealm in shiro with
  spring}}.
\newblock
  \bibinfo{howpublished}{\url{https://stackoverflow.com/questions/30463057/authc-filter-is-not-calling-myrealm-in-shiro-with-spring}}.
    (\bibinfo{year}{[n. d.]}).
\newblock


\bibitem[\protect\citeauthoryear{??}{key}{[n. d.]b}]%
        {keyfromkeystore}
 \bibinfo{year}{[n. d.]}\natexlab{b}.
\newblock \bibinfo{title}{{Basic Program for encrypt/Decrypt :
  javax.crypto.BadPaddingException: Decryption error}}.
\newblock
  \bibinfo{howpublished}{\url{https://stackoverflow.com/questions/39518979/basic-program-for-encrypt-decrypt-javax-crypto-badpaddingexception-decryption}}.
    (\bibinfo{year}{[n. d.]}).
\newblock


\bibitem[\protect\citeauthoryear{??}{key}{[n. d.]c}]%
        {keyfrombigintegers}
 \bibinfo{year}{[n. d.]}\natexlab{c}.
\newblock \bibinfo{title}{{BigInteger to Key}}.
\newblock
  \bibinfo{howpublished}{\url{https://stackoverflow.com/questions/10271164/biginteger-to-key}}.
    (\bibinfo{year}{[n. d.]}).
\newblock


\bibitem[\protect\citeauthoryear{??}{bc}{[n. d.]}]%
        {bc}
 \bibinfo{year}{[n. d.]}\natexlab{}.
\newblock \bibinfo{title}{Bouncy Castle}.
\newblock \bibinfo{howpublished}{\url{https://www.bouncycastle.org}}.
  (\bibinfo{year}{[n. d.]}).
\newblock


\bibitem[\protect\citeauthoryear{??}{ref}{[n. d.]}]%
        {reflectionenvironment}
 \bibinfo{year}{[n. d.]}\natexlab{}.
\newblock \bibinfo{title}{{Can a secret be hidden in a 'safe' java class
  offering access credentials?}}
\newblock
  \bibinfo{howpublished}{\url{https://stackoverflow.com/questions/5761519/can-a-secret-be-hidden-in-a-safe-java-class-offering-access-credentials}}.
    (\bibinfo{year}{[n. d.]}).
\newblock


\bibitem[\protect\citeauthoryear{??}{key}{[n. d.]d}]%
        {keyforcomparison}
 \bibinfo{year}{[n. d.]}\natexlab{d}.
\newblock \bibinfo{title}{{Compare two Public Key values in java [duplicate]}}.
\newblock
  \bibinfo{howpublished}{\url{https://stackoverflow.com/questions/37439695/compare-two-public-key-values-in-java}}.
    (\bibinfo{year}{[n. d.]}).
\newblock


\bibitem[\protect\citeauthoryear{??}{xml}{[n. d.]}]%
        {xmltojava}
 \bibinfo{year}{[n. d.]}\natexlab{}.
\newblock \bibinfo{title}{{Configure Spring Security without XML in Spring 4}}.
\newblock
  \bibinfo{howpublished}{\url{https://stackoverflow.com/questions/20961600/configure-spring-security-without-xml-in-spring-4}}.
    (\bibinfo{year}{[n. d.]}).
\newblock


\bibitem[\protect\citeauthoryear{??}{con}{[n. d.]a}]%
        {conflictannotationcode}
 \bibinfo{year}{[n. d.]}\natexlab{a}.
\newblock \bibinfo{title}{{@Context injection in Stateless EJB used by
  JAX-RS}}.
\newblock
  \bibinfo{howpublished}{\url{https://stackoverflow.com/questions/29132547/context-injection-in-stateless-ejb-used-by-jax-rs}}.
    (\bibinfo{year}{[n. d.]}).
\newblock


\bibitem[\protect\citeauthoryear{??}{key}{[n. d.]e}]%
        {keyfrombytes}
 \bibinfo{year}{[n. d.]}\natexlab{e}.
\newblock \bibinfo{title}{{Converted secret key into bytes, how to convert it
  back to secrect key?}}
\newblock
  \bibinfo{howpublished}{\url{https://stackoverflow.com/questions/5364338/converted-secret-key-into-bytes-how-to-convert-it-back-to-secrect-key}}.
    (\bibinfo{year}{[n. d.]}).
\newblock


\bibitem[\protect\citeauthoryear{??}{cwe}{[n. d.]}]%
        {cwe227}
 \bibinfo{year}{[n. d.]}\natexlab{}.
\newblock \bibinfo{title}{CWE-227: Improper Fulfillment of API Contract ('API
  Abuse')}.
\newblock
  \bibinfo{howpublished}{\url{https://cwe.mitre.org/data/definitions/227.html}}.
    (\bibinfo{year}{[n. d.]}).
\newblock


\bibitem[\protect\citeauthoryear{??}{dic}{[n. d.]}]%
        {dictionaryattack}
 \bibinfo{year}{[n. d.]}\natexlab{}.
\newblock \bibinfo{title}{{Dictionary Attacks 101}}.
\newblock
  \bibinfo{howpublished}{\url{https://blog.codinghorror.com/dictionary-attacks-101/}}.
    (\bibinfo{year}{[n. d.]}).
\newblock


\bibitem[\protect\citeauthoryear{??}{key}{[n. d.]f}]%
        {keyfromstring}
 \bibinfo{year}{[n. d.]}\natexlab{f}.
\newblock \bibinfo{title}{{Edit code sample to specify DES key value}}.
\newblock \bibinfo{howpublished}{\url{Edit code sample to specify DES key
  value}}.   (\bibinfo{year}{[n. d.]}).
\newblock


\bibitem[\protect\citeauthoryear{??}{com}{[n. d.]}]%
        {commonencryptionerror}
 \bibinfo{year}{[n. d.]}\natexlab{}.
\newblock \bibinfo{title}{{Encryption PHP, Decryption Java}}.
\newblock
  \bibinfo{howpublished}{\url{https://stackoverflow.com/questions/15639442/encryption-php-decryption-java}}.
    (\bibinfo{year}{[n. d.]}).
\newblock


\bibitem[\protect\citeauthoryear{??}{key}{[n. d.]g}]%
        {keyfromcertificate}
 \bibinfo{year}{[n. d.]}\natexlab{g}.
\newblock \bibinfo{title}{{Get public and private key from ASN1 encrypted pem
  certificate}}.
\newblock
  \bibinfo{howpublished}{https://stackoverflow.com/questions/30392114/get-public-and-private-key-from-asn1-encrypted-pem-certificate}.
    (\bibinfo{year}{[n. d.]}).
\newblock


\bibitem[\protect\citeauthoryear{??}{gla}{[n. d.]}]%
        {glassfish}
 \bibinfo{year}{[n. d.]}\natexlab{}.
\newblock \bibinfo{title}{{GlassFish}}.
\newblock \bibinfo{howpublished}{\url{https://javaee.github.io/glassfish/}}.
  (\bibinfo{year}{[n. d.]}).
\newblock


\bibitem[\protect\citeauthoryear{??}{nis}{[n. d.]}]%
        {nist}
 \bibinfo{year}{[n. d.]}\natexlab{}.
\newblock \bibinfo{title}{{Guidelines for the Selection, Configuration, and Use
  of Transport Layer Security (TLS) Implementations}}.
\newblock
  \bibinfo{howpublished}{\url{http://nvlpubs.nist.gov/nistpubs/SpecialPublications/NIST.SP.800-52r1.pdf}}.
    (\bibinfo{year}{[n. d.]}).
\newblock


\bibitem[\protect\citeauthoryear{??}{lim}{[n. d.]a}]%
        {limitmemberaccess}
 \bibinfo{year}{[n. d.]}\natexlab{a}.
\newblock \bibinfo{title}{{Hiding my security key from java reflection}}.
\newblock
  \bibinfo{howpublished}{\url{https://stackoverflow.com/questions/14903318/hiding-my-security-key-from-java-reflection}}.
    (\bibinfo{year}{[n. d.]}).
\newblock


\bibitem[\protect\citeauthoryear{??}{app}{[n. d.]}]%
        {appletjs}
 \bibinfo{year}{[n. d.]}\natexlab{}.
\newblock \bibinfo{title}{{How can I get a signed Java Applet to perform
  privileged operations when called from unsigned Javascript?}}
\newblock
  \bibinfo{howpublished}{\url{https://stackoverflow.com/questions/1006674/how-can-i-get-a-signed-java-applet-to-perform-privileged-operations-when-called}}.
    (\bibinfo{year}{[n. d.]}).
\newblock


\bibitem[\protect\citeauthoryear{??}{152}{[n. d.]}]%
        {15274874}
 \bibinfo{year}{[n. d.]}\natexlab{}.
\newblock \bibinfo{title}{{How does Java string being immutable increase
  security?}}
\newblock
  \bibinfo{howpublished}{\url{https://stackoverflow.com/questions/15274874/how-does-java-string-being-immutable-increase-security}}.
    (\bibinfo{year}{[n. d.]}).
\newblock


\bibitem[\protect\citeauthoryear{??}{ssl}{[n. d.]a}]%
        {ssljndi}
 \bibinfo{year}{[n. d.]}\natexlab{a}.
\newblock \bibinfo{title}{{how to accept self-signed certificates for JNDI/LDAP
  connections?}}
\newblock
  \bibinfo{howpublished}{\url{https://stackoverflow.com/questions/4615163/how-to-accept-self-signed-certificates-for-jndi-ldap-connections}}.
    (\bibinfo{year}{[n. d.]}).
\newblock


\bibitem[\protect\citeauthoryear{??}{rep}{[n. d.]}]%
        {reputation}
 \bibinfo{year}{[n. d.]}\natexlab{}.
\newblock \bibinfo{title}{{How to add MD5 or SHA hash to spring security?}}
\newblock
  \bibinfo{howpublished}{\url{https://stackoverflow.com/questions/18581463/how-to-add-md5-or-sha-hash-to-spring-security}}.
    (\bibinfo{year}{[n. d.]}).
\newblock


\bibitem[\protect\citeauthoryear{??}{sec}{[n. d.]a}]%
        {securitymethodorder}
 \bibinfo{year}{[n. d.]}\natexlab{a}.
\newblock \bibinfo{title}{{How to apply spring security filter only on secured
  endpoints?}}
\newblock
  \bibinfo{howpublished}{\url{https://stackoverflow.com/questions/36795894/how-to-apply-spring-security-filter-only-on-secured-endpoints}}.
    (\bibinfo{year}{[n. d.]}).
\newblock


\bibitem[\protect\citeauthoryear{??}{key}{[n. d.]h}]%
        {keyfromnumber}
 \bibinfo{year}{[n. d.]}\natexlab{h}.
\newblock \bibinfo{title}{{How to generate secret key using
  SecureRandom.getInstanceStrong()?}}
\newblock
  \bibinfo{howpublished}{\url{https://stackoverflow.com/questions/37244064/how-to-generate-secret-key-using-securerandom-getinstancestrong}}.
    (\bibinfo{year}{[n. d.]}).
\newblock


\bibitem[\protect\citeauthoryear{??}{spr}{[n. d.]a}]%
        {springjava}
 \bibinfo{year}{[n. d.]}\natexlab{a}.
\newblock \bibinfo{title}{{How to override Spring Security default
  configuration in Spring Boot}}.
\newblock
  \bibinfo{howpublished}{\url{https://stackoverflow.com/questions/35600488/how-to-override-spring-security-default-configuration-in-spring-boot}}.
    (\bibinfo{year}{[n. d.]}).
\newblock


\bibitem[\protect\citeauthoryear{??}{spr}{[n. d.]b}]%
        {springsecuritydoc}
 \bibinfo{year}{[n. d.]}\natexlab{b}.
\newblock \bibinfo{title}{{HttpSecurity (Spring Security 4.2.3.RELEASE API)}}.
\newblock
  \bibinfo{howpublished}{\url{https://docs.spring.io/spring-security/site/docs/current/apidocs/org/springframework/security/config/annotation/web/builders/HttpSecurity.html}}.
    (\bibinfo{year}{[n. d.]}).
\newblock


\bibitem[\protect\citeauthoryear{??}{rmi}{[n. d.]a}]%
        {rmitutorial}
 \bibinfo{year}{[n. d.]}\natexlab{a}.
\newblock \bibinfo{title}{{Implementing a Remote Interface}}.
\newblock
  \bibinfo{howpublished}{\url{http://docs.oracle.com/javase/tutorial/rmi/implementing.html}}.
    (\bibinfo{year}{[n. d.]}).
\newblock


\bibitem[\protect\citeauthoryear{??}{pkc}{[n. d.]}]%
        {pkcs8}
 \bibinfo{year}{[n. d.]}\natexlab{}.
\newblock \bibinfo{title}{{InvalidKeySpecException : algid parse error, not a
  sequence}}.
\newblock
  \bibinfo{howpublished}{\url{https://stackoverflow.com/questions/31941413/invalidkeyspecexception-algid-parse-error-not-a-sequence}}.
    (\bibinfo{year}{[n. d.]}).
\newblock


\bibitem[\protect\citeauthoryear{??}{csr}{[n. d.]a}]%
        {csrftoken}
 \bibinfo{year}{[n. d.]}\natexlab{a}.
\newblock \bibinfo{title}{{java - Simple example of Spring Security with
  Thymeleaf}}.
\newblock
  \bibinfo{howpublished}{\url{https://stackoverflow.com/questions/25692735/simple-example-of-spring-security-with-thymeleaf}}.
    (\bibinfo{year}{[n. d.]}).
\newblock


\bibitem[\protect\citeauthoryear{??}{jaa}{[n. d.]}]%
        {jaasref}
 \bibinfo{year}{[n. d.]}\natexlab{}.
\newblock \bibinfo{title}{Java Authentication and Authorization Service (JAAS)
  Reference Guide}.
\newblock
  \bibinfo{howpublished}{\url{https://docs.oracle.com/javase/8/docs/technotes/guides/security/jaas/JAASRefGuide.html}}.
    (\bibinfo{year}{[n. d.]}).
\newblock


\bibitem[\protect\citeauthoryear{??}{pos}{[n. d.]}]%
        {post}
 \bibinfo{year}{[n. d.]}\natexlab{}.
\newblock \bibinfo{title}{{java class to trust all for sending file to https
  web service}}.
\newblock
  \bibinfo{howpublished}{\url{https://stackoverflow.com/questions/21156929/java-class-to-trust-all-for-sending-file-to-https-web-service}}.
    (\bibinfo{year}{[n. d.]}).
\newblock


\bibitem[\protect\citeauthoryear{??}{jca}{[n. d.]}]%
        {jca}
 \bibinfo{year}{[n. d.]}\natexlab{}.
\newblock \bibinfo{title}{Java Cryptography Architecture}.
\newblock
  \bibinfo{howpublished}{\url{http://docs.oracle.com/javase/7/docs/technotes/guides/security/crypto/CryptoSpec.html}}.
    (\bibinfo{year}{[n. d.]}).
\newblock


\bibitem[\protect\citeauthoryear{??}{con}{[n. d.]b}]%
        {conflictannotations}
 \bibinfo{year}{[n. d.]}\natexlab{b}.
\newblock \bibinfo{title}{{Java EE 7 EJB Security not working}}.
\newblock
  \bibinfo{howpublished}{\url{https://stackoverflow.com/questions/30504131/java-ee-7-ejb-security-not-working}}.
    (\bibinfo{year}{[n. d.]}).
\newblock


\bibitem[\protect\citeauthoryear{??}{est}{[n. d.]}]%
        {establishssl}
 \bibinfo{year}{[n. d.]}\natexlab{}.
\newblock \bibinfo{title}{{Java Mail get mails with pop3 from exchange server
  => Exception in thread ``main'' javax.mail.MessagingException}}.
\newblock
  \bibinfo{howpublished}{\url{https://stackoverflow.com/questions/25017050/java-mail-get-mails-with-pop3-from-exchange-server-exception-in-thread-main}}.
    (\bibinfo{year}{[n. d.]}).
\newblock


\bibitem[\protect\citeauthoryear{??}{rmi}{[n. d.]b}]%
        {rmism}
 \bibinfo{year}{[n. d.]}\natexlab{b}.
\newblock \bibinfo{title}{{Java RMI / access denied}}.
\newblock
  \bibinfo{howpublished}{\url{https://stackoverflow.com/questions/36570012/java-rmi-access-denied}}.
    (\bibinfo{year}{[n. d.]}).
\newblock


\bibitem[\protect\citeauthoryear{??}{key}{[n. d.]i}]%
        {keyfromscratch}
 \bibinfo{year}{[n. d.]}\natexlab{i}.
\newblock \bibinfo{title}{{Java Security - RSA Public Key \& Private Key Code
  Issue}}.
\newblock
  \bibinfo{howpublished}{\url{https://stackoverflow.com/questions/18757114/java-security-rsa-public-key-private-key-code-issue}}.
    (\bibinfo{year}{[n. d.]}).
\newblock


\bibitem[\protect\citeauthoryear{??}{key}{[n. d.]j}]%
        {keyforcipher}
 \bibinfo{year}{[n. d.]}\natexlab{j}.
\newblock \bibinfo{title}{{Java security init Cipher from SecretKeySpec
  properly}}.
\newblock
  \bibinfo{howpublished}{\url{https://stackoverflow.com/questions/14230096/java-security-init-cipher-from-secretkeyspec-properly}}.
    (\bibinfo{year}{[n. d.]}).
\newblock


\bibitem[\protect\citeauthoryear{??}{lim}{[n. d.]b}]%
        {limitpackageaccess}
 \bibinfo{year}{[n. d.]}\natexlab{b}.
\newblock \bibinfo{title}{{Java Security Manager completely disable
  reflection}}.
\newblock
  \bibinfo{howpublished}{\url{https://stackoverflow.com/questions/40218973/java-security-manager-completely-disable-reflection}}.
    (\bibinfo{year}{[n. d.]}).
\newblock


\bibitem[\protect\citeauthoryear{??}{jSe}{[n. d.]}]%
        {jSecurity}
 \bibinfo{year}{[n. d.]}\natexlab{}.
\newblock \bibinfo{title}{Java Security Overview}.
\newblock
  \bibinfo{howpublished}{\url{http://docs.oracle.com/javase/8/docs/technotes/guides/security/overview/jsoverview.html}}.
    (\bibinfo{year}{[n. d.]}).
\newblock


\bibitem[\protect\citeauthoryear{??}{lim}{[n. d.]c}]%
        {limitclassaccess}
 \bibinfo{year}{[n. d.]}\natexlab{c}.
\newblock \bibinfo{title}{{Java security: Sandboxing plugins loaded via
  URLClassLoader}}.
\newblock
  \bibinfo{howpublished}{\url{https://stackoverflow.com/questions/3947558/java-security-sandboxing-plugins-loaded-via-urlclassloader}}.
    (\bibinfo{year}{[n. d.]}).
\newblock


\bibitem[\protect\citeauthoryear{??}{ssl}{[n. d.]b}]%
        {sslcertificate}
 \bibinfo{year}{[n. d.]}\natexlab{b}.
\newblock \bibinfo{title}{{Java SSL - InstallCert recognizes certificate, but
  still ``unable to find valid certification path'' error?}}
\newblock
  \bibinfo{howpublished}{\url{https://stackoverflow.com/questions/11087121/java-ssl-installcert-recognizes-certificate-but-still-unable-to-find-valid-c}}.
    (\bibinfo{year}{[n. d.]}).
\newblock


\bibitem[\protect\citeauthoryear{??}{jee}{[n. d.]}]%
        {jeesecurity}
 \bibinfo{year}{[n. d.]}\natexlab{}.
\newblock \bibinfo{title}{JSR-000366 Java Platform, Enterprise Edition 8 Public
  Review Specification}.
\newblock
  \bibinfo{howpublished}{\url{http://download.oracle.com/otndocs/jcp/java_ee-8-pr-spec/}}.
    (\bibinfo{year}{[n. d.]}).
\newblock


\bibitem[\protect\citeauthoryear{??}{csr}{[n. d.]b}]%
        {csrfdisable}
 \bibinfo{year}{[n. d.]}\natexlab{b}.
\newblock \bibinfo{title}{{logout call - Spring security logout call}}.
\newblock
  \bibinfo{howpublished}{\url{https://stackoverflow.com/questions/24530603/spring-security-logout-call}}.
    (\bibinfo{year}{[n. d.]}).
\newblock


\bibitem[\protect\citeauthoryear{??}{md5}{[n. d.]}]%
        {md5}
 \bibinfo{year}{[n. d.]}\natexlab{}.
\newblock \bibinfo{title}{{MD5 hashing in Android}}.
\newblock
  \bibinfo{howpublished}{\url{https://stackoverflow.com/questions/4846484/md5-hashing-in-android}}.
    (\bibinfo{year}{[n. d.]}).
\newblock


\bibitem[\protect\citeauthoryear{??}{con}{[n. d.]c}]%
        {conflictannotationpath}
 \bibinfo{year}{[n. d.]}\natexlab{c}.
\newblock \bibinfo{title}{{PicketLink / Deltaspike security does not work in
  SOAP (JAX-WS) layer (CDI vs EJB?)}}.
\newblock
  \bibinfo{howpublished}{\url{https://stackoverflow.com/questions/32392702/picketlink-deltaspike-security-does-not-work-in-soap-jax-ws-layer-cdi-vs-ej}}.
    (\bibinfo{year}{[n. d.]}).
\newblock


\bibitem[\protect\citeauthoryear{??}{jav}{[n. d.]}]%
        {javaeecombination}
 \bibinfo{year}{[n. d.]}\natexlab{}.
\newblock \bibinfo{title}{{Resteasy Authorization design - check a user owns a
  resource}}.
\newblock
  \bibinfo{howpublished}{\url{https://stackoverflow.com/questions/34315838/resteasy-authorization-design-check-a-user-owns-a-resource}}.
    (\bibinfo{year}{[n. d.]}).
\newblock


\bibitem[\protect\citeauthoryear{??}{rfs}{[n. d.]}]%
        {rfs6101}
 \bibinfo{year}{[n. d.]}\natexlab{}.
\newblock \bibinfo{title}{{RF 6101 - The Secure Sockets Layer (SSL) Protocol
  Version 3.0}}.
\newblock \bibinfo{howpublished}{\url{https://tools.ietf.org/html/rfc6101}}.
  (\bibinfo{year}{[n. d.]}).
\newblock


\bibitem[\protect\citeauthoryear{??}{scr}{[n. d.]}]%
        {scrapy}
 \bibinfo{year}{[n. d.]}\natexlab{}.
\newblock \bibinfo{title}{{Scrapy | A Fast and Powerful Scraping and Web
  Crawling Framework}}.
\newblock \bibinfo{howpublished}{\url{https://scrapy.org}}.
  (\bibinfo{year}{[n. d.]}).
\newblock


\bibitem[\protect\citeauthoryear{??}{ssl}{[n. d.]c}]%
        {ssltypical}
 \bibinfo{year}{[n. d.]}\natexlab{c}.
\newblock \bibinfo{title}{{security - Allowing Java to use an untrusted
  certificate for SSL/HTTPS connection}}.
\newblock
  \bibinfo{howpublished}{\url{https://stackoverflow.com/questions/1201048/allowing-java-to-use-an-untrusted-certificate-for-ssl-https-connection}}.
    (\bibinfo{year}{[n. d.]}).
\newblock


\bibitem[\protect\citeauthoryear{??}{all}{[n. d.]}]%
        {allowapplet}
 \bibinfo{year}{[n. d.]}\natexlab{}.
\newblock \bibinfo{title}{{security exception when loading web image in jar}}.
\newblock
  \bibinfo{howpublished}{\url{https://stackoverflow.com/questions/2011407/security-exception-when-loading-web-image-in-jar}}.
    (\bibinfo{year}{[n. d.]}).
\newblock


\bibitem[\protect\citeauthoryear{??}{spr}{[n. d.]c}]%
        {springsecurity}
 \bibinfo{year}{[n. d.]}\natexlab{c}.
\newblock \bibinfo{title}{Spring Security}.
\newblock
  \bibinfo{howpublished}{\url{https://projects.spring.io/spring-security/}}.
  (\bibinfo{year}{[n. d.]}).
\newblock


\bibitem[\protect\citeauthoryear{??}{sec}{[n. d.]b}]%
        {securityspringmvc}
 \bibinfo{year}{[n. d.]}\natexlab{b}.
\newblock \bibinfo{title}{{Spring Security 4 xml configuration
  UserDetailsService authentication not working}}.
\newblock
  \bibinfo{howpublished}{\url{https://stackoverflow.com/questions/41321176/spring-security-4-xml-configuration-userdetailsservice-authentication-not-workin}}.
    (\bibinfo{year}{[n. d.]}).
\newblock


\bibitem[\protect\citeauthoryear{??}{spr}{[n. d.]d}]%
        {springxml}
 \bibinfo{year}{[n. d.]}\natexlab{d}.
\newblock \bibinfo{title}{{Spring security JDK based proxy issue while using
  @Secured annotation on Controller method}}.
\newblock
  \bibinfo{howpublished}{\url{https://stackoverflow.com/questions/35860442/spring-security-jdk-based-proxy-issue-while-using-secured-annotation-on-control}}.
    (\bibinfo{year}{[n. d.]}).
\newblock


\bibitem[\protect\citeauthoryear{??}{spr}{[n. d.]e}]%
        {springsecurityreference32}
 \bibinfo{year}{[n. d.]}\natexlab{e}.
\newblock \bibinfo{title}{{Spring Security Reference}}.
\newblock
  \bibinfo{howpublished}{\url{http://docs.spring.io/spring-security/site/docs/3.2.4.RELEASE/reference/htmlsingle/\#jc-httpsecurity}}.
    (\bibinfo{year}{[n. d.]}).
\newblock


\bibitem[\protect\citeauthoryear{??}{spr}{[n. d.]f}]%
        {springvariousapplications}
 \bibinfo{year}{[n. d.]}\natexlab{f}.
\newblock \bibinfo{title}{{Spring Security Tutorial}}.
\newblock
  \bibinfo{howpublished}{\url{http://www.mkyong.com/tutorials/spring-security-tutorials/}}.
    (\bibinfo{year}{[n. d.]}).
\newblock


\bibitem[\protect\citeauthoryear{??}{spr}{[n. d.]g}]%
        {springjboss}
 \bibinfo{year}{[n. d.]}\natexlab{g}.
\newblock \bibinfo{title}{{Spring Security using JBoss <security-domain>}}.
\newblock
  \bibinfo{howpublished}{\url{https://stackoverflow.com/questions/28172056/spring-security-using-jboss-security-domain}}.
    (\bibinfo{year}{[n. d.]}).
\newblock


\bibitem[\protect\citeauthoryear{??}{cer}{[n. d.]a}]%
        {certcreate}
 \bibinfo{year}{[n. d.]}\natexlab{a}.
\newblock \bibinfo{title}{{SSL Certificate Verification :
  javax.net.ssl.SSLHandshakeException}}.
\newblock
  \bibinfo{howpublished}{\url{https://stackoverflow.com/questions/25079751/ssl-certificate-verification-javax-net-ssl-sslhandshakeexception}}.
    (\bibinfo{year}{[n. d.]}).
\newblock


\bibitem[\protect\citeauthoryear{??}{ssl}{[n. d.]d}]%
        {sslworkaround}
 \bibinfo{year}{[n. d.]}\natexlab{d}.
\newblock \bibinfo{title}{{Ssl handshake fails with unable to find valid
  certification path to requested target}}.
\newblock
  \bibinfo{howpublished}{\url{https://stackoverflow.com/questions/40977556/ssl-handshake-fails-with-unable-to-find-valid-certification-path-to-requested-ta}}.
    (\bibinfo{year}{[n. d.]}).
\newblock


\bibitem[\protect\citeauthoryear{??}{cer}{[n. d.]b}]%
        {certuse}
 \bibinfo{year}{[n. d.]}\natexlab{b}.
\newblock \bibinfo{title}{{SSL Socket Connection working even though client is
  not sending certificate?}}
\newblock
  \bibinfo{howpublished}{\url{https://stackoverflow.com/questions/26761966/ssl-socket-connection-working-even-though-client-is-not-sending-certificate}}.
    (\bibinfo{year}{[n. d.]}).
\newblock


\bibitem[\protect\citeauthoryear{??}{sta}{[n. d.]}]%
        {stackoverflow}
 \bibinfo{year}{[n. d.]}\natexlab{}.
\newblock \bibinfo{title}{StackOverflow}.
\newblock \bibinfo{howpublished}{\url{https://stackoverflow.com}}.
  (\bibinfo{year}{[n. d.]}).
\newblock


\bibitem[\protect\citeauthoryear{??}{ver}{[n. d.]}]%
        {veracode}
 \bibinfo{year}{[n. d.]}\natexlab{}.
\newblock \bibinfo{title}{STATE OF SOFTWARE SECURITY}.
\newblock
  \bibinfo{howpublished}{{https://www.veracode.com/sites/default/files/Resources/Reports/state-of-software-security-volume-7-veracode-report.pdf}}.
    (\bibinfo{year}{[n. d.]}).
\newblock


\bibitem[\protect\citeauthoryear{??}{cer}{[n. d.]c}]%
        {certinstall}
 \bibinfo{year}{[n. d.]}\natexlab{c}.
\newblock \bibinfo{title}{{The Webserver I talk to updated its SSL cert and now
  my app can't talk to it}}.
\newblock
  \bibinfo{howpublished}{\url{https://stackoverflow.com/questions/5758812/the-webserver-i-talk-to-updated-its-ssl-cert-and-now-my-app-cant-talk-to-it}}.
    (\bibinfo{year}{[n. d.]}).
\newblock


\bibitem[\protect\citeauthoryear{??}{ssl}{[n. d.]e}]%
        {ssldebate}
 \bibinfo{year}{[n. d.]}\natexlab{e}.
\newblock \bibinfo{title}{{Trusting all certificates using HttpClient over
  HTTPS}}.
\newblock
  \bibinfo{howpublished}{\url{https://stackoverflow.com/questions/2642777/trusting-all-certificates-using-httpclient-over-https}}.
    (\bibinfo{year}{[n. d.]}).
\newblock


\bibitem[\protect\citeauthoryear{??}{key}{[n. d.]k}]%
        {keyforprint}
 \bibinfo{year}{[n. d.]}\natexlab{k}.
\newblock \bibinfo{title}{{Use of ECC in Java SE 1.7}}.
\newblock
  \bibinfo{howpublished}{\url{https://stackoverflow.com/questions/24383637/use-of-ecc-in-java-se-1-7}}.
    (\bibinfo{year}{[n. d.]}).
\newblock


\bibitem[\protect\citeauthoryear{??}{key}{[n. d.]l}]%
        {keyfromfile}
 \bibinfo{year}{[n. d.]}\natexlab{l}.
\newblock \bibinfo{title}{{Using public key from authorized\_keys with Java
  security}}.
\newblock
  \bibinfo{howpublished}{\url{https://stackoverflow.com/questions/3531506/using-public-key-from-authorized-keys-with-java-security}}.
    (\bibinfo{year}{[n. d.]}).
\newblock


\bibitem[\protect\citeauthoryear{??}{spr}{[n. d.]h}]%
        {springsample}
 \bibinfo{year}{[n. d.]}\natexlab{h}.
\newblock \bibinfo{title}{{Web Security Samples}}.
\newblock
  \bibinfo{howpublished}{\url{https://github.com/spring-projects/spring-security-javaconfig/blob/master/samples-web.md\#sample-multi-http-web-configuration}}.
    (\bibinfo{year}{[n. d.]}).
\newblock


\bibitem[\protect\citeauthoryear{??}{web}{[n. d.]b}]%
        {websphere}
 \bibinfo{year}{[n. d.]}\natexlab{b}.
\newblock \bibinfo{title}{{WebSphere Application Server - IBM}}.
\newblock
  \bibinfo{howpublished}{\url{http://www-03.ibm.com/software/products/en/appserv-was}}.
    (\bibinfo{year}{[n. d.]}).
\newblock


\bibitem[\protect\citeauthoryear{??}{ssl}{[n. d.]f}]%
        {sslpowermock}
 \bibinfo{year}{[n. d.]}\natexlab{f}.
\newblock \bibinfo{title}{{When a TrustManagerFactory is not a
  TrustManagerFactory (Java)}}.
\newblock
  \bibinfo{howpublished}{\url{https://stackoverflow.com/questions/14654639/when-a-trustmanagerfactory-is-not-a-trustmanagerfactory-java}}.
    (\bibinfo{year}{[n. d.]}).
\newblock


\bibitem[\protect\citeauthoryear{??}{wil}{[n. d.]}]%
        {wildfly}
 \bibinfo{year}{[n. d.]}\natexlab{}.
\newblock \bibinfo{title}{{WildFly}}.
\newblock \bibinfo{howpublished}{\url{http://wildfly.org}}.
  (\bibinfo{year}{[n. d.]}).
\newblock


\bibitem[\protect\citeauthoryear{??}{con}{[n. d.]d}]%
        {conflictannotationdescriptor}
 \bibinfo{year}{[n. d.]}\natexlab{d}.
\newblock \bibinfo{title}{{Wildfly 9 security domains won't work}}.
\newblock
  \bibinfo{howpublished}{\url{https://stackoverflow.com/questions/37425056/wildfly-9-security-domains-wont-work}}.
    (\bibinfo{year}{[n. d.]}).
\newblock


\bibitem[\protect\citeauthoryear{??}{412}{[n. d.]}]%
        {41278592}
 \bibinfo{year}{[n. d.]}\natexlab{}.
\newblock \bibinfo{title}{{Your implementation of PreferenceActivity is
  vulnerable to fragment injection}}.
\newblock
  \bibinfo{howpublished}{\url{https://stackoverflow.com/questions/41278592/your-implementation-of-preferenceactivity-is-vulnerable-to-fragment-injection}}.
    (\bibinfo{year}{[n. d.]}).
\newblock


\bibitem[\protect\citeauthoryear{Boonkrong}{Boonkrong}{2012}]%
        {boonkrong2012security}
\bibfield{author}{\bibinfo{person}{Sirapat Boonkrong}.}
  \bibinfo{year}{2012}\natexlab{}.
\newblock \showarticletitle{Security of passwords}.
\newblock \bibinfo{journal}{{\em Information Technology Journal\/}}
  \bibinfo{volume}{8}, \bibinfo{number}{2} (\bibinfo{year}{2012}),
  \bibinfo{pages}{112--117}.
\newblock


\bibitem[\protect\citeauthoryear{Chatzikonstantinou, Ntantogian, Karopoulos,
  and Xenakis}{Chatzikonstantinou et~al\mbox{.}}{2016}]%
        {Chatzikonstantinou:2016}
\bibfield{author}{\bibinfo{person}{Alexia Chatzikonstantinou},
  \bibinfo{person}{Christoforos Ntantogian}, \bibinfo{person}{Georgios
  Karopoulos}, {and} \bibinfo{person}{Christos Xenakis}.}
  \bibinfo{year}{2016}\natexlab{}.
\newblock \showarticletitle{Evaluation of Cryptography Usage in Android
  Applications}. In \bibinfo{booktitle}{{\em Proceedings of the 9th EAI
  International Conference on Bio-inspired Information and Communications
  Technologies (Formerly BIONETICS)}} {\em (\bibinfo{series}{BICT'15})}.
  \bibinfo{publisher}{ICST (Institute for Computer Sciences, Social-Informatics
  and Telecommunications Engineering)}, \bibinfo{address}{ICST, Brussels,
  Belgium, Belgium}, \bibinfo{pages}{83--90}.
\newblock
\showISBNx{978-1-63190-100-3}
\showDOI{%
\url{https://doi.org/10.4108/eai.3-12-2015.2262471}}


\bibitem[\protect\citeauthoryear{Datta, Derek, Mitchell, and Roy}{Datta
  et~al\mbox{.}}{2007}]%
        {Datta2007}
\bibfield{author}{\bibinfo{person}{Anupam Datta}, \bibinfo{person}{Ante Derek},
  \bibinfo{person}{John~C. Mitchell}, {and} \bibinfo{person}{Arnab Roy}.}
  \bibinfo{year}{2007}\natexlab{}.
\newblock \showarticletitle{Protocol Composition Logic (PCL)}.
\newblock \bibinfo{journal}{{\em Electronic Notes in Theoretical Computer
  Science\/}}  \bibinfo{volume}{172} (\bibinfo{year}{2007}),
  \bibinfo{pages}{311 -- 358}.
\newblock
\showISSN{1571-0661}
\showDOI{%
\url{https://doi.org/10.1016/j.entcs.2007.02.012}}
\newblock
\shownote{Computation, Meaning, and Logic: Articles dedicated to Gordon
  Plotkin.}


\bibitem[\protect\citeauthoryear{Dey and Weis}{Dey and Weis}{[n. d.]}]%
        {keyczar}
\bibfield{author}{\bibinfo{person}{Arkajit Dey} {and} \bibinfo{person}{Stephen
  Weis}.} \bibinfo{year}{[n. d.]}\natexlab{}.
\newblock \bibinfo{booktitle}{{\em Keyczar: A Cryptographic Toolkit}}.
\newblock


\bibitem[\protect\citeauthoryear{Egele, Brumley, Fratantonio, and
  Kruegel}{Egele et~al\mbox{.}}{2013}]%
        {Egele:2013}
\bibfield{author}{\bibinfo{person}{Manuel Egele}, \bibinfo{person}{David
  Brumley}, \bibinfo{person}{Yanick Fratantonio}, {and}
  \bibinfo{person}{Christopher Kruegel}.} \bibinfo{year}{2013}\natexlab{}.
\newblock \showarticletitle{An Empirical Study of Cryptographic Misuse in
  Android Applications}. In \bibinfo{booktitle}{{\em Proceedings of the 2013
  ACM SIGSAC Conference on Computer \&\#38; Communications Security}} {\em
  (\bibinfo{series}{CCS '13})}. \bibinfo{publisher}{ACM}, \bibinfo{address}{New
  York, NY, USA}, \bibinfo{pages}{73--84}.
\newblock
\showISBNx{978-1-4503-2477-9}
\showDOI{%
\url{https://doi.org/10.1145/2508859.2516693}}


\bibitem[\protect\citeauthoryear{Erk\"{o}k and Matthews}{Erk\"{o}k and
  Matthews}{2008}]%
        {Erkok:2009}
\bibfield{author}{\bibinfo{person}{Levent Erk\"{o}k} {and}
  \bibinfo{person}{John Matthews}.} \bibinfo{year}{2008}\natexlab{}.
\newblock \showarticletitle{Pragmatic Equivalence and Safety Checking in
  Cryptol}. In \bibinfo{booktitle}{{\em Proceedings of the 3rd Workshop on
  Programming Languages Meets Program Verification}} {\em
  (\bibinfo{series}{PLPV '09})}. \bibinfo{publisher}{ACM},
  \bibinfo{address}{New York, NY, USA}, \bibinfo{pages}{73--82}.
\newblock
\showISBNx{978-1-60558-330-3}
\showDOI{%
\url{https://doi.org/10.1145/1481848.1481860}}


\bibitem[\protect\citeauthoryear{Fahl, Harbach, Muders, Baumg\"{a}rtner,
  Freisleben, and Smith}{Fahl et~al\mbox{.}}{2012}]%
        {Fahl:2012}
\bibfield{author}{\bibinfo{person}{Sascha Fahl}, \bibinfo{person}{Marian
  Harbach}, \bibinfo{person}{Thomas Muders}, \bibinfo{person}{Lars
  Baumg\"{a}rtner}, \bibinfo{person}{Bernd Freisleben}, {and}
  \bibinfo{person}{Matthew Smith}.} \bibinfo{year}{2012}\natexlab{}.
\newblock \showarticletitle{Why Eve and Mallory Love Android: An Analysis of
  Android SSL (in)Security}. In \bibinfo{booktitle}{{\em Proceedings of the
  2012 ACM Conference on Computer and Communications Security}} {\em
  (\bibinfo{series}{CCS '12})}. \bibinfo{publisher}{ACM}, \bibinfo{address}{New
  York, NY, USA}, \bibinfo{pages}{50--61}.
\newblock
\showISBNx{978-1-4503-1651-4}
\showDOI{%
\url{https://doi.org/10.1145/2382196.2382205}}


\bibitem[\protect\citeauthoryear{Fischer, B¨ottinger, Xiao, Stransky, Acar,
  Backes, and Fahl}{Fischer et~al\mbox{.}}{2017}]%
        {fischer:2017}
\bibfield{author}{\bibinfo{person}{Felix Fischer}, \bibinfo{person}{Konstantin
  B¨ottinger}, \bibinfo{person}{Huang Xiao}, \bibinfo{person}{Christian
  Stransky}, \bibinfo{person}{Yasemin Acar}, \bibinfo{person}{Michael Backes},
  {and} \bibinfo{person}{Sascha Fahl}.} \bibinfo{year}{2017}\natexlab{}.
\newblock \showarticletitle{Stack Overflow Considered Harmful? The Impact of
  Copy\&Paste on Android Application Security}. In \bibinfo{booktitle}{{\em
  38th IEEE Symposium on Security and Privacy (S\&P '17)}} (2017-05-22).
\newblock


\bibitem[\protect\citeauthoryear{Gackenheimer}{Gackenheimer}{2013}]%
        {gackenheimer2013implementing}
\bibfield{author}{\bibinfo{person}{Cory Gackenheimer}.}
  \bibinfo{year}{2013}\natexlab{}.
\newblock \showarticletitle{Implementing Security and Cryptography}.
\newblock In \bibinfo{booktitle}{{\em Node. js Recipes}}.
  \bibinfo{publisher}{Springer}, \bibinfo{pages}{133--160}.
\newblock


\bibitem[\protect\citeauthoryear{Georgiev, Iyengar, Jana, Anubhai, Boneh, and
  Shmatikov}{Georgiev et~al\mbox{.}}{2012}]%
        {Georgiev:2012}
\bibfield{author}{\bibinfo{person}{Martin Georgiev}, \bibinfo{person}{Subodh
  Iyengar}, \bibinfo{person}{Suman Jana}, \bibinfo{person}{Rishita Anubhai},
  \bibinfo{person}{Dan Boneh}, {and} \bibinfo{person}{Vitaly Shmatikov}.}
  \bibinfo{year}{2012}\natexlab{}.
\newblock \showarticletitle{The Most Dangerous Code in the World: Validating
  SSL Certificates in Non-browser Software}. In \bibinfo{booktitle}{{\em
  Proceedings of the 2012 ACM Conference on Computer and Communications
  Security}} {\em (\bibinfo{series}{CCS '12})}. \bibinfo{publisher}{ACM},
  \bibinfo{address}{New York, NY, USA}, \bibinfo{pages}{38--49}.
\newblock
\showISBNx{978-1-4503-1651-4}
\showDOI{%
\url{https://doi.org/10.1145/2382196.2382204}}


\bibitem[\protect\citeauthoryear{Gong and Ellison}{Gong and Ellison}{2003}]%
        {Gong:2003}
\bibfield{author}{\bibinfo{person}{Li Gong} {and} \bibinfo{person}{Gary
  Ellison}.} \bibinfo{year}{2003}\natexlab{}.
\newblock \bibinfo{booktitle}{{\em Inside Java(TM) 2 Platform Security:
  Architecture, API Design, and Implementation\/} (\bibinfo{edition}{2nd}
  ed.)}.
\newblock \bibinfo{publisher}{Pearson Education}.
\newblock
\showISBNx{0201787911}


\bibitem[\protect\citeauthoryear{He, Rastogi, Cao, Chen, Venkatakrishnan, Yang,
  and Zhang}{He et~al\mbox{.}}{2015}]%
        {He2015}
\bibfield{author}{\bibinfo{person}{B. He}, \bibinfo{person}{V. Rastogi},
  \bibinfo{person}{Y. Cao}, \bibinfo{person}{Y. Chen}, \bibinfo{person}{V.~N.
  Venkatakrishnan}, \bibinfo{person}{R. Yang}, {and} \bibinfo{person}{Z.
  Zhang}.} \bibinfo{year}{2015}\natexlab{}.
\newblock \showarticletitle{Vetting SSL Usage in Applications with SSLINT}. In
  \bibinfo{booktitle}{{\em 2015 IEEE Symposium on Security and Privacy}}.
  \bibinfo{pages}{519--534}.
\newblock
\showISSN{1081-6011}
\showDOI{%
\url{https://doi.org/10.1109/SP.2015.38}}


\bibitem[\protect\citeauthoryear{Lazar, Chen, Wang, and Zeldovich}{Lazar
  et~al\mbox{.}}{2014}]%
        {Lazar:2014}
\bibfield{author}{\bibinfo{person}{David Lazar}, \bibinfo{person}{Haogang
  Chen}, \bibinfo{person}{Xi Wang}, {and} \bibinfo{person}{Nickolai
  Zeldovich}.} \bibinfo{year}{2014}\natexlab{}.
\newblock \showarticletitle{Why Does Cryptographic Software Fail?: A Case Study
  and Open Problems}. In \bibinfo{booktitle}{{\em Proceedings of 5th
  Asia-Pacific Workshop on Systems}} {\em (\bibinfo{series}{APSys '14})}.
  \bibinfo{publisher}{ACM}, \bibinfo{address}{New York, NY, USA}, Article
  \bibinfo{articleno}{7}, \bibinfo{numpages}{7}~pages.
\newblock
\showISBNx{978-1-4503-3024-4}
\showDOI{%
\url{https://doi.org/10.1145/2637166.2637237}}


\bibitem[\protect\citeauthoryear{Li, Zhang, Li, and Gu}{Li
  et~al\mbox{.}}{2014}]%
        {Li2014}
\bibfield{author}{\bibinfo{person}{Yong Li}, \bibinfo{person}{Yuanyuan Zhang},
  \bibinfo{person}{Juanru Li}, {and} \bibinfo{person}{Dawu Gu}.}
  \bibinfo{year}{2014}\natexlab{}.
\newblock \bibinfo{booktitle}{{\em iCryptoTracer: Dynamic Analysis on Misuse of
  Cryptography Functions in iOS Applications}}.
\newblock \bibinfo{publisher}{Springer International Publishing},
  \bibinfo{address}{Cham}, \bibinfo{pages}{349--362}.
\newblock
\showISBNx{978-3-319-11698-3}
\showDOI{%
\url{https://doi.org/10.1007/978-3-319-11698-3_27}}


\bibitem[\protect\citeauthoryear{Long}{Long}{2005}]%
        {LongSoftwareVulnerabilities2005}
\bibfield{author}{\bibinfo{person}{Fred Long}.}
  \bibinfo{year}{2005}\natexlab{}.
\newblock \bibinfo{booktitle}{{\em Software Vulnerabilities in Java}}.
\newblock \bibinfo{type}{{T}echnical {R}eport} CMU/SEI-2005-TN-044.
  \bibinfo{institution}{Software Engineering Institute, Carnegie Mellon
  University}, \bibinfo{address}{Pittsburgh, PA}.
\newblock
\showURL{%
\url{http://resources.sei.cmu.edu/library/asset-view.cfm?AssetID=7573}}


\bibitem[\protect\citeauthoryear{Mettler, Wagner, and Close}{Mettler
  et~al\mbox{.}}{2010}]%
        {MettlerWagnerClose10}
\bibfield{author}{\bibinfo{person}{Adrian Mettler}, \bibinfo{person}{David
  Wagner}, {and} \bibinfo{person}{Tyler Close}.}
  \bibinfo{year}{2010}\natexlab{}.
\newblock \showarticletitle{Joe-E: A Security-Oriented Subset of Java}. In
  \bibinfo{booktitle}{{\em Network and Distributed Systems Symposium}}.
  Internet Society.
\newblock
\showURL{%
\url{http://www.truststc.org/pubs/652.html}}


\bibitem[\protect\citeauthoryear{Mitchell, Mitchell, and Stern}{Mitchell
  et~al\mbox{.}}{1997}]%
        {Mitchell:1997}
\bibfield{author}{\bibinfo{person}{J.~C. Mitchell}, \bibinfo{person}{M.
  Mitchell}, {and} \bibinfo{person}{U. Stern}.}
  \bibinfo{year}{1997}\natexlab{}.
\newblock \showarticletitle{Automated Analysis of Cryptographic Protocols Using
  Mur/Spl Phi/}. In \bibinfo{booktitle}{{\em Proceedings of the 1997 IEEE
  Symposium on Security and Privacy}} {\em (\bibinfo{series}{SP '97})}.
  \bibinfo{publisher}{IEEE Computer Society}, \bibinfo{address}{Washington, DC,
  USA}, \bibinfo{pages}{141--}.
\newblock
\showURL{%
\url{http://dl.acm.org/citation.cfm?id=882493.884384}}


\bibitem[\protect\citeauthoryear{M{\"o}ller, Duong, and Kotowicz}{M{\"o}ller
  et~al\mbox{.}}{2014}]%
        {moller2014poodle}
\bibfield{author}{\bibinfo{person}{Bodo M{\"o}ller}, \bibinfo{person}{Thai
  Duong}, {and} \bibinfo{person}{Krzysztof Kotowicz}.}
  \bibinfo{year}{2014}\natexlab{}.
\newblock \showarticletitle{This POODLE bites: exploiting the SSL 3.0
  fallback}.
\newblock \bibinfo{journal}{{\em PDF online\/}} (\bibinfo{year}{2014}),
  \bibinfo{pages}{1--4}.
\newblock


\bibitem[\protect\citeauthoryear{Nadi, Kr\"{u}ger, Mezini, and Bodden}{Nadi
  et~al\mbox{.}}{2016}]%
        {Nadi:2016}
\bibfield{author}{\bibinfo{person}{Sarah Nadi}, \bibinfo{person}{Stefan
  Kr\"{u}ger}, \bibinfo{person}{Mira Mezini}, {and} \bibinfo{person}{Eric
  Bodden}.} \bibinfo{year}{2016}\natexlab{}.
\newblock \showarticletitle{Jumping Through Hoops: Why Do Java Developers
  Struggle with Cryptography APIs?}. In \bibinfo{booktitle}{{\em Proceedings of
  the 38th International Conference on Software Engineering}} {\em
  (\bibinfo{series}{ICSE '16})}. \bibinfo{publisher}{ACM},
  \bibinfo{address}{New York, NY, USA}, \bibinfo{pages}{935--946}.
\newblock
\showISBNx{978-1-4503-3900-1}
\showDOI{%
\url{https://doi.org/10.1145/2884781.2884790}}


\bibitem[\protect\citeauthoryear{Oaks}{Oaks}{1998}]%
        {Oaks:1998}
\bibfield{author}{\bibinfo{person}{Scott Oaks}.}
  \bibinfo{year}{1998}\natexlab{}.
\newblock \bibinfo{booktitle}{{\em Java Security}}.
\newblock \bibinfo{publisher}{O'Reilly \& Associates, Inc.},
  \bibinfo{address}{Sebastopol, CA, USA}.
\newblock
\showISBNx{1-56592-403-7}


\bibitem[\protect\citeauthoryear{Onwuzurike and De~Cristofaro}{Onwuzurike and
  De~Cristofaro}{2015}]%
        {Onwuzurike:2015}
\bibfield{author}{\bibinfo{person}{Lucky Onwuzurike} {and}
  \bibinfo{person}{Emiliano De~Cristofaro}.} \bibinfo{year}{2015}\natexlab{}.
\newblock \showarticletitle{Danger is My Middle Name: Experimenting with SSL
  Vulnerabilities in Android Apps}. In \bibinfo{booktitle}{{\em Proceedings of
  the 8th ACM Conference on Security \& Privacy in Wireless and Mobile
  Networks}} {\em (\bibinfo{series}{WiSec '15})}. \bibinfo{publisher}{ACM},
  \bibinfo{address}{New York, NY, USA}, Article \bibinfo{articleno}{15},
  \bibinfo{numpages}{6}~pages.
\newblock
\showISBNx{978-1-4503-3623-9}
\showDOI{%
\url{https://doi.org/10.1145/2766498.2766522}}


\bibitem[\protect\citeauthoryear{Rashid}{Rashid}{[n. d.]}]%
        {libmisuse}
\bibfield{author}{\bibinfo{person}{Fahmida~Y. Rashid}.} \bibinfo{year}{[n.
  d.]}\natexlab{}.
\newblock \bibinfo{title}{Library misuse exposes leading Java platforms to
  attack}.
\newblock
  \bibinfo{howpublished}{\url{http://www.infoworld.com/article/3003197/security/library-misuse-exposes-leading-java-platforms-to-attack.html}}.
    (\bibinfo{year}{[n. d.]}).
\newblock


\bibitem[\protect\citeauthoryear{Shuai, Guowei, Tao, Tianchang, and
  Chenjie}{Shuai et~al\mbox{.}}{2014}]%
        {Shuai:2014}
\bibfield{author}{\bibinfo{person}{Shao Shuai}, \bibinfo{person}{Dong Guowei},
  \bibinfo{person}{Guo Tao}, \bibinfo{person}{Yang Tianchang}, {and}
  \bibinfo{person}{Shi Chenjie}.} \bibinfo{year}{2014}\natexlab{}.
\newblock \showarticletitle{Modelling Analysis and Auto-detection of
  Cryptographic Misuse in Android Applications}. In \bibinfo{booktitle}{{\em
  Proceedings of the 2014 IEEE 12th International Conference on Dependable,
  Autonomic and Secure Computing}} {\em (\bibinfo{series}{DASC '14})}.
  \bibinfo{publisher}{IEEE Computer Society}, \bibinfo{address}{Washington, DC,
  USA}, \bibinfo{pages}{75--80}.
\newblock
\showISBNx{978-1-4799-5079-9}
\showDOI{%
\url{https://doi.org/10.1109/DASC.2014.22}}


\bibitem[\protect\citeauthoryear{Smith and Dill}{Smith and Dill}{2008}]%
        {Smith2008}
\bibfield{author}{\bibinfo{person}{E. Smith} {and} \bibinfo{person}{D.~L.
  Dill}.} \bibinfo{year}{2008}\natexlab{}.
\newblock \showarticletitle{Automatic Formal Verification of Block Cipher
  Implementations}. In \bibinfo{booktitle}{{\em 2008 Formal Methods in
  Computer-Aided Design}}. \bibinfo{pages}{1--7}.
\newblock
\showDOI{%
\url{https://doi.org/10.1109/FMCAD.2008.ECP.10}}


\bibitem[\protect\citeauthoryear{Steven}{Steven}{2017}]%
        {owaspcheatsheet17}
\bibfield{author}{\bibinfo{person}{J Steven}.} \bibinfo{year}{2017}\natexlab{}.
\newblock \showarticletitle{Password storage cheat sheet}.
\newblock  (\bibinfo{year}{2017}).
\newblock
\showURL{%
\url{https://www.owasp.org/index.php/Password_Storage_Cheat_Sheet}}


\bibitem[\protect\citeauthoryear{Stevens, Bursztein, Karpman, Albertini, and
  Markov}{Stevens et~al\mbox{.}}{2017}]%
        {sha1stevens17}
\bibfield{author}{\bibinfo{person}{Marc Stevens}, \bibinfo{person}{Elie
  Bursztein}, \bibinfo{person}{Pierre Karpman}, \bibinfo{person}{Ange
  Albertini}, {and} \bibinfo{person}{Yarik Markov}.}
  \bibinfo{year}{2017}\natexlab{}.
\newblock \showarticletitle{The first collision for full SHA-1.}
\newblock \bibinfo{journal}{{\em IACR Cryptology ePrint Archive\/}}
  \bibinfo{volume}{2017} (\bibinfo{year}{2017}), \bibinfo{pages}{190}.
\newblock


\bibitem[\protect\citeauthoryear{Wang, Feng, Lai, and Yu}{Wang
  et~al\mbox{.}}{2004}]%
        {md5wang04}
\bibfield{author}{\bibinfo{person}{Xiaoyun Wang}, \bibinfo{person}{Dengguo
  Feng}, \bibinfo{person}{Xuejia Lai}, {and} \bibinfo{person}{Hongbo Yu}.}
  \bibinfo{year}{2004}\natexlab{}.
\newblock \showarticletitle{Collisions for Hash Functions MD4, MD5, HAVAL-128
  and RIPEMD.}
\newblock \bibinfo{journal}{{\em IACR Cryptology ePrint Archive\/}}
  \bibinfo{volume}{2004} (\bibinfo{year}{2004}), \bibinfo{pages}{199}.
\newblock


\bibitem[\protect\citeauthoryear{Yang, Lo, Xia, Wan, and Sun}{Yang
  et~al\mbox{.}}{2016}]%
        {Yang2016}
\bibfield{author}{\bibinfo{person}{Xin-Li Yang}, \bibinfo{person}{David Lo},
  \bibinfo{person}{Xin Xia}, \bibinfo{person}{Zhi-Yuan Wan}, {and}
  \bibinfo{person}{Jian-Ling Sun}.} \bibinfo{year}{2016}\natexlab{}.
\newblock \showarticletitle{What Security Questions Do Developers Ask? A
  Large-Scale Study of Stack Overflow Posts}.
\newblock \bibinfo{journal}{{\em Journal of Computer Science and Technology\/}}
  \bibinfo{volume}{31}, \bibinfo{number}{5} (\bibinfo{date}{01 Sep}
  \bibinfo{year}{2016}), \bibinfo{pages}{910--924}.
\newblock
\showISSN{1860-4749}
\showDOI{%
\url{https://doi.org/10.1007/s11390-016-1672-0}}


\bibitem[\protect\citeauthoryear{Zeller and Felten}{Zeller and Felten}{[n.
  d.]}]%
        {zellercross}
\bibfield{author}{\bibinfo{person}{William Zeller} {and}
  \bibinfo{person}{Edward~W Felten}.} \bibinfo{year}{[n. d.]}\natexlab{}.
\newblock \bibinfo{title}{Cross-Site Request Forgeries: Exploitation and
  prevention, 2008}.
\newblock   (\bibinfo{year}{[n. d.]}).
\newblock


\end{thebibliography}

\end{document}